\documentclass{article}
\pdfoutput=1
\usepackage{amsmath,amssymb,graphicx,microtype,hyperref}
\usepackage{amsthm}

\usepackage{framed}

\usepackage{lscape}

\def\be{\begin{eqnarray}}
\def\ee{\end{eqnarray}}
\def\nn{\nonumber}

\def\p{\partial}

\def\l[{\phantom.[}

\newcommand{\tket}[1]{| #1 \rangle }

\newcommand{\NPb}{{\rlap{\raise.4ex\hbox{$\scriptscriptstyle\bullet$}}
\lower.4ex\hbox{$\scriptscriptstyle\bullet$}}}
\newcommand{\NPc}{{\rlap{\raise.4ex\hbox{$\scriptscriptstyle\circ$}}
\lower.4ex\hbox{$\scriptscriptstyle\circ$}}}

\newcommand{\vl}{\vec{\lambda}}
\newcommand{\vm}{\vec{\mu}}

\newcommand{\lo}{\lambda^{(1)}}

\newcommand{\lN}{\lambda^{(N)}}

\newcommand{\vu}{\vec{u}}

\newcommand{\Xo}{X^{(1)}}

\newcommand{\tM}{\widetilde{M}}

\newcommand{\po}{p^{(1)}}
\newcommand{\pt}{p^{(2)}}

\theoremstyle{definition}
\newtheorem{df}{Definition}[section]

\newtheorem{conj}[df]{Conjecture}

\hoffset= - 1.0in         % switch off for draft style
\begin{document}

\title{\vspace{.1cm}{\Large {\bf Toric Calabi-Yau threefolds as
      quantum integrable systems.\\
      $\mathcal{R}$-matrix and $\mathcal{RTT}$ relations
      % Ding-Iohara-Miki ${\cal R}$-matrix and
% knot hyper???polynomials
}\vspace{.2cm}}
\author{
{\bf Hidetoshi Awata$^a$}\footnote{awata@math.nagoya-u.ac.jp},
\ {\bf Hiroaki Kanno$^{a,b}$}\footnote{kanno@math.nagoya-u.ac.jp},
\ {\bf Andrei Mironov$^{c,d,e,f}$}\footnote{mironov@lpi.ru; mironov@itep.ru},
\ {\bf Alexei Morozov$^{d,e,f}$}\thanks{morozov@itep.ru},\\
\ {\bf Andrey Morozov$^{d,e,f,g}$}\footnote{andrey.morozov@itep.ru},
\ {\bf Yusuke Ohkubo$^a$}\footnote{m12010t@math.nagoya-u.ac.jp}
\ \ and \ {\bf Yegor Zenkevich$^{d,f,h}$}\thanks{yegor.zenkevich@gmail.com}}
\date{ }
}

\maketitle

\vspace{-6.5cm}

\begin{center}
\hfill FIAN/TD-20/16\\
\hfill IITP/TH-15/16\\
\hfill ITEP/TH-21/16\\
\hfill INR-TH-2016-30
\end{center}

\vspace{4.3cm}

\begin{center}
$^a$ {\small {\it Graduate School of Mathematics, Nagoya University,
Nagoya, 464-8602, Japan}}\\
$^b$ {\small {\it KMI, Nagoya University,
Nagoya, 464-8602, Japan}}\\
$^c$ {\small {\it Lebedev Physics Institute, Moscow 119991, Russia}}\\
$^d$ {\small {\it ITEP, Moscow 117218, Russia}}\\
$^e$ {\small {\it Institute for Information Transmission Problems, Moscow 127994, Russia}}\\
$^f$ {\small {\it National Research Nuclear University MEPhI, Moscow 115409, Russia }}\\
$^g$ {\small {\it Laboratory of Quantum Topology, Chelyabinsk State University, Chelyabinsk 454001, Russia }}\\
$^h$ {\small {\it Institute of Nuclear Research, Moscow 117312, Russia }}
\end{center}

\vspace{.5cm}

\begin{abstract}
${\cal R}$-matrix is explicitly constructed for simplest
  representations of the Ding-Iohara-Miki algebra. Calculation is
  straightforward and significantly simpler than the one through the
  universal $\mathcal{R}$-matrix used for a similar calculation in the Yangian case by
  A.~Smirnov but less general. We investigate the interplay
  between the $\mathcal{R}$-matrix structure and the structure of DIM
  algebra intertwiners, i.e.\ of refined topological vertices and show
  that the $\mathcal{R}$-matrix is diagonalized by the action of the
  spectral duality belonging to the $SL(2,\mathbb{Z})$ group of DIM
  algebra automorphisms. We also construct the $\mathcal{T}$-operators
  satisfying the $\mathcal{RTT}$ relations with
  the $\mathcal{R}$-matrix from refined amplitudes on resolved
  conifold. We thus show that topological string theories on the toric Calabi-Yau threefolds can be
  naturally interpreted as lattice integrable models. Integrals of
  motion for these systems are related to $q$-deformation of the
  reflection matrices of the Liouville/Toda theories.
\end{abstract}

\vspace{.5cm}

\section{Introduction }
\label{sec:introduction-}
Integrability plays an exceptional role in modern studies of quantum
field theory and string theory. Whenever there is a breakthrough in understanding of
non-perturbative dynamics, some form of
integrability invariably appears to be behind this success. An
(incomplete) list of recent examples includes
\begin{itemize}
\item Seiberg-Witten solution of $\mathcal{N}=2$ theories \cite{SW} and the
  corresponding classical complex integrable systems \cite{SWint}

\item integrability in $\mathcal{N}=4$ gauge theory and the AdS/CFT
  dual string theory, coming from integrable spin chains and
  $\sigma$-models \cite{N4int}

\item Seiberg dualities in $\mathcal{N}=1$ gauge theories \cite{Sdual} and the
  corresponding integrable lattice models with new solutions to
  Yang-Baxter equations \cite{N1int}

\item AGT relations \cite{AGT}, integrability \cite{AGTint} and new family of integrals of motion in
  $W_N$-algebras related to the basis of fixed points in the instanton
  moduli space \cite{ILW}

\item topological string calculations and the study of Hurwitz
  $\tau$-functions \cite{Hurwitz}
\end{itemize}
In this paper we demonstrate a new integrable structure in
refined topological strings on toric Calabi-Yau threefolds. This
structure is related to several points from the list above and we
elaborate on these connections in sec.~\ref{sec:conclusion}. Let us
now briefly summarize how this kind of integrability appears.

The central object, on which we will mostly focus in our approach is
the $\mathcal{R}$-matrix of the Ding-Iohara-Miki (DIM) algebra \cite{DI,Miki}.
${\cal R}$-matrices, which can be considered as emerging in the description of
coproducts of group elements $\hat g\in {\cal G}\otimes {{\cal A(G)}}$ \cite{qge}
for quantum groups \cite{JD},
\begin{equation}
  (I\otimes \hat g)\cdot (\hat g\otimes I) =
  {\cal R}\cdot (\hat g\otimes I)\cdot (I\otimes \hat g)\cdot {\cal R}^{-1},\label{eq:17}
\end{equation}
are crucial to all integrable systems. As evident from
Eq.~(\ref{eq:17}), the job of the $\mathcal{R}$-matrix is to permute
the components in the tensor product of representations of the algebra
$\mathcal{G}$. This is the property we will use in refined topological
strings. The representations in question are going to be Fock modules \cite{DIMreps}
and their permutation exchanges the legs of the toric diagram
corresponding to a DIM intertwiner \cite{AFS,DIM1}.

The permutation of the legs performed by the $\mathcal{R}$-matrix has
a simple interpretation in terms of the corresponding conformal blocks
of the $q$-Virasoro or $qW_N$-algebras. Ratios of the spectral
parameters on the horizontal legs determine the Liouville-like momenta
of the primary states \cite{DIM1}. By exchanging the spectral parameters, the
$\mathcal{R}$-matrix inverts the momenta, and therefore acts exactly
as the Liouville reflection matrix introduced in~\cite{ZZ}. This
connection (first noted in \cite{MaOk}, see also \cite{M1}) is quite interesting, since, as we will see in the
following, the $\mathcal{R}$-matrix can be evaluated explicitly by
solving for the eigenfunctions of the generalized Macdonald
Hamiltonian with known eigenvalues.

Also among other things, let us mention that the $\mathcal{R}$-matrices
are used to construct knot polynomials in Chern-Simons theory
\cite{CS}, one of the most challenging subjects in topology.  In
particular, the knot superpolynomials of \cite{Super}, constructed
with the help of double-affine Hecke algebras (DAHA) \cite{Che}, still lack a
clear ${\cal R}$-matrix realization within the Reshetikhin-Turaev (RT)
formalism, either original \cite{RT} or modern \cite{RTmod}. On the other hand, the DIM algebra
is naturally related with DAHA by a kind of Schur duality (see \cite{VV} for a degenerate version of this correspondence). There is another way to naturally associate these two algebras: the DIM algebra is the limit of spherical DAHA for large number of
strands (see \cite{SV} for a degenerate version of this correspondence).

The notation in this paper follows our paper \cite{DIM1}.

\subsection{DIM algebra, generalized Macdonald polynomials and the
  $\mathcal{R}$-matrix}
\label{sec:dim-algebra-gener}
We are going to compute the $\mathcal{R}$-matrix of the DIM algebra,
also known as quantum toroidal algebra or
$U_{q,t}(\widehat{\widehat{\mathfrak{gl}}}_1)$ \cite{DI,Miki}. It is a double quantum
deformation of the double loop algebra of $\mathfrak{gl}_1$. The double
loop algebra can be understood as the algebra of torus mappings into
the group $\mathfrak{gl}_{1}$. The two deformation parameters are
related to the quantum deformation parameter of the affine algebra
$\widehat{\mathfrak{gl}}_1$ and the quantum deformation of the torus
respectively.

\begin{figure}[h]
  \centering
  \includegraphics[width=8cm]{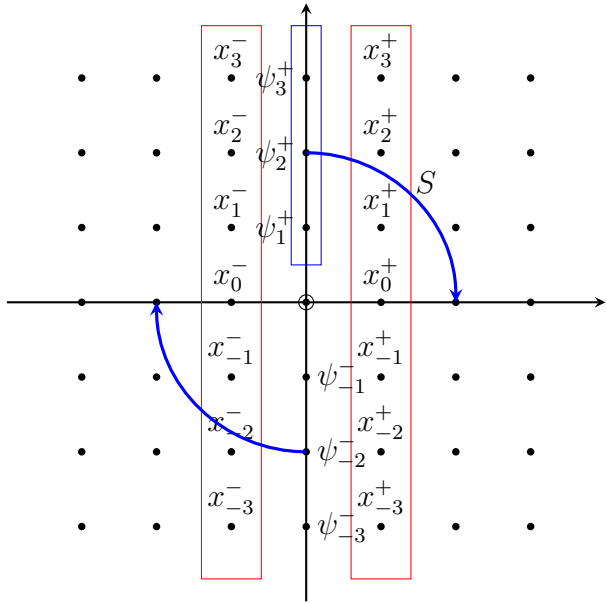}
  \caption{The lattice of
    $U_{q,t}(\widehat{\widehat{\mathfrak{gl}}}_1)$ generators. The
    algebra is doubly graded, so that each generator has two integer
    weights. The standard generators $x^{+}_n$, $\psi^{\pm}_n$ and
    $x^{-}_n$ form the three central rows. $\psi^{+}_n$ generators
    (they are framed in blue) form the Cartan subalgebra and $x^{+}_n$, $x^{-}_n$
    act as the raising and lowering operators respectively. Blue
    arrows show the action $\mathcal{S}$ of the spectral duality, an
    $SL(2,\mathbb{Z})$ rotation of the integer lattice. Notice that
    $\mathcal{S}(\psi^{+}_1) = x^{+}_0$, i.e.\ the first Cartan generator
    transforms into the zero mode of the raising generator.}
  \label{fig:1}
\end{figure}

The DIM algebra is generated by respectively the ``raising'' and ``lowering''
operators $x^{+}_n$ and $x^{-}_n$ with $n \in \mathbb{Z}$
together with the ``Cartan'' generators $\psi^{\pm}_{\pm n}$, $n \in \mathbb{Z}_{>0}$ and two central
elements $C_1$, $C_2$. The algebra has a double grading coming from the two
loops, i.e. a torus $T^2$, in the double loop construction. Each
element of the algebra with a definite grading can be, therefore, drawn
as an integral point on the plane. The generators, $x^{+}_n$, $\psi^{\pm}_n$,
$x^{-}_n$ and their commutators form a \emph{lattice}, which is
sketched in Fig.~\ref{fig:1}. The exact definition of the DIM algebra can be found in \cite{DIM,awata2011notes,DIM1} (see also \cite{ellDIM} for elliptic DIM algebra).

There is a nice representation of the DIM algebra on the Fock space
$\mathcal{F}_u^{(1,0)}$, i.e.\ a \emph{bosonization} of the DIM
generators, which are expressed through exponentials of the free bosons
(for concrete expressions see \cite{DIMreps,AFS}). The second central charge
of this representation is trivial, $C_2=1$, while the first one is
given by $C_1 = (t/q)^{1/2}$. We will henceforth call this
representation \emph{horizontal}, since the first central charge is
associated with the horizontal direction. There is also the
\emph{vertical} Fock representation $\mathcal{F}^{(0,1)}_u$,
isomorphic to the horizontal one, but with a different action of the
DIM generators \cite{DIMreps,AFS}. In the basis of Macdonald symmetric polynomials,
$M_Y^{(q,t)}(a_{-n}) | u \rangle$, the generators $x^{+}_n$ add a box
to the Young diagram $Y$, while $x^{-}_n$ delete one box, and
$\psi^{\pm}_n$ act diagonally. The central charges of this
representation are $(1, (t/q)^{1/2})$ (we refer the reader to
\cite{DIMreps,AFS} for the complete construction).

It will be important for us that the DIM algebra has a remarkable
group of automorphisms $SL(2,\mathbb{Z})$, which are precisely the
automorphisms of the integer lattice of generators \cite{Miki}. Let us also note
that the central charges $(C_1, C_2)$ transform as a doublet under
this $SL(2,\mathbb{Z})$ symmetry. One of the automorphisms, which we
call $\mathcal{S}$ is particularly important\footnote{$\mathcal{S}$ is
  sometimes called \emph{Miki isomorphism} in the mathematical
  literature. In physical terms, it is Type IIB $S$-duality exchanging
  NS5 and D5 branes, hence, our notation.}. $\mathcal{S}$ corresponds
to rotation of the integer lattice by $\frac{\pi}{2}$ clockwise. The
action of this element on the algebra realizes the \emph{spectral
  duality} \cite{specdu,triality} of different representations: in particular, the central
charge vector is rotated; the horizontal representations become the vertical ones
and vice versa. The action of $\mathcal{S}$ is illustrated in
Fig.~\ref{fig:1}.

Let us construct a natural basis in the tensor product of horizontal
modules. This basis is given by \emph{generalized} Macdonald
polynomials \cite{awata2011notes,MoSmi,genMac,FJMM-Bethe,Z}
$\widetilde{M}_{AB}\left(\frac{u_1}{u_2}\Big| q,t\Big| p_n^{(1)}, p_n^{(2)}
\right)$, which are the eigenfunctions
\begin{equation}
  {\cal H}_1\, \widetilde{M}_{AB} = \kappa_{AB} \widetilde{M}_{AB}
\end{equation}
of the Hamiltonian
\begin{equation}
 {\cal H}_1 = \oint \frac{dz}{z} \
\rho_{u_1}\otimes \rho_{u_2}\left\{ \Delta_{\rm DIM}(x^{+}(z))\right\}
\label{Ham}
\end{equation}
with eigenvalues
\begin{equation}
  \label{eq:18}
  \kappa_{AB} = u_1 \sum_{i\geq 1} q^{A_i} t^{-i} + u_2 \sum_{i\geq 1}
  q^{B_i} t^{-i}.
\end{equation}
In the simplest example, i.e.\ for the tensor product of \emph{two}
Fock modules $\mathcal{F}_{u_1} \otimes \mathcal{F}_{u_2}$, the
generalized Macdonald polynomials depend on a pair of Young diagrams
and on ratio of the spectral parameters $\frac{u_1}{u_2}$.

The Hamiltonian $\hat{\mathcal{H}}_1$ is the zero mode of the raising
generator, $x^{+}_0$ in the horizontal representation. One can also
understand the Hamiltonian~\eqref{Ham} as the \emph{spectral dual} of
the first Cartan generator $\psi^{+}_1$. As we mentioned above, in the
vertical representation the Cartan generators $\psi^{+}_n$ acts diagonally
on the \emph{ordinary} Macdonald polynomials. The same is true for tensor
products of the vertical representations, i.e.\ the ``diagonal'' basis is
given by tensor products of the Macdonald polynomials
$M_A^{(q,t)}(a_{-n}^{(1)}) |u_1 \rangle \otimes
M_B^{(q,t)}(a_{-n}^{(2)}) |u_2 \rangle$ (in order to see this, one should use
the DIM coproduct \cite{DIM1} and the fact that, for the vertical representations,
$C_1=1$). Thus, the \emph{generalized} Macdonald polynomials
$\widetilde{M}_{AB}\left( \frac{u_1}{u_2} \Big| q,t \Big|a_{-n}^{(1)},
  a_{-n}^{(2)}\right) |u_1 \rangle \otimes |u_2 \rangle$, which
diagonalize $x^{+}_0 = \mathcal{S}(\psi^{+}_1)$, can be thought of as
spectral duals of the ordinary Macdonald
polynomials.
A remarkable feature of DIM, which greatly simplifies calculations, is
that the eigenvalues of the first Hamiltonian $\mathcal{H}_1$
are non-degenerate, so it is sufficient to diagonalize only this one
operator to define the entire set of polynomials and all ``higher
Hamiltonians'' (i.e.\ the other Cartan generators, $\mathcal{H}_n =
\rho_{u_1} \otimes \rho_{u_2} \mathcal{S}(\psi^{+}_n)$ for $n \geq 2$) are
automatically diagonal, see Appendix B.

Let us make two remarks here. The eigenvalues are non-degenerate only
for $u_1$, $u_2$ in \emph{general position}. However, the case of
\emph{resonance} between $u_1$ and $u_2$ is more subtle, then the
eigenvalues do become degenerate. We will not consider this case. The
eigenvalues also become degenerate in the $4d$/Yangian limit in which
the first Hamiltonian should be expanded up to the first order and a lot of
information is thus lost. This is because the $(q,t)$-deformation reveals
a true exponential nature of the DIM-symmetry generators, while the ordinary
Virasoro and ${\cal W}$ (and thus the higher Hamiltonians of the
Calogero-Sutherland-Ruijsennars family) arise all together in their
series expansions. We will usually suppress the two sets of
time variables $p_n=\sum_i x_i^n$, $\bar p_n=\sum_i \bar x_i^n$,
which the Hamiltonian acts on and polynomials depend on; when they are
needed, we use the notation $M\{p,\bar p\}$ or $M[x,\bar x]$,
depending on the choice between the time and Miwa parametrizations.

In the tensor products of more than two Fock modules, there are still
eigenstates of $x^{+}_0$, which we call in the same way
generalized Macdonald polynomials. In this case, the number of time
sets and Young diagrams is correspondingly increased.

Now we are at the crucial point of our approach to the
$\mathcal{R}$-matrix. The Hamiltonian $\hat {\cal H}_1$ depends on
the choice of the coproduct in the DIM algebra; there are two natural
options: schematically,
\begin{equation}
\Delta(x^{+}) = x^{+}\otimes 1 + \psi^{-} \otimes x^{+}
\end{equation}
or
\begin{equation}
\Delta^{\mathrm{op}}(x^{+}) =1\otimes x^{+} + x^{+} \otimes \psi^{-}.
\end{equation}
The DIM algebra is a quasitriangular Hopf algebra. Thus, these two coproducts are related by an ${\cal R}$-matrix:
\begin{equation}
  \Delta^{\mathrm{op}} = {\cal R}\Delta{\cal R}^{-1}, \qquad \hat{\cal H}^{\mathrm{op}}_1 = {\cal R}  \hat{{\cal H}} {\cal R}^{-1}
\end{equation}
Hence, their eigenfunctions are also related\footnote{Notice a slight
  change in the notation compared to~\cite{Z}. We are now writing the
  generalized Macdonald polynomials as functions of the variable
  $\frac{u_1}{u_2}$, which we call $Q$, whereas in~\cite{Z} we
  denoted $\frac{u_2}{u_1}$ as $Q$.}:
\begin{equation}
  \widetilde{M}_{AB}^{\mathrm{op}}\left(\frac{u_1}{u_2}\Big| q,t \Big| p,\bar p\right) =
  \sum_{C,D} {\cal R}^{CD}_{AB}\!\left(\frac{u_1}{u_2}\right)\cdot
  \widetilde{M}_{CD}\left(\frac{u_1}{u_2}\Big|q,t \Big| p, \bar p\right)
\label{tildeMvsM}
\end{equation}
where the sum is actually finite, because the size of Young diagrams
is restricted by the conservation law
\be
|A|+|B| = |C|+|D|
\ee
which makes ${\cal R}$ block-diagonal with finite-dimensional blocks.

The coproducts $\Delta$ and $\Delta^{\mathrm{op}}$ differ only by permutation of the two representations on which the algebra
acts. Thus, the ``opposite'' Macdonald polynomials can be
alternatively obtained by a simple change of variables, exchanging
$u_1 \leftrightarrow u_2$, $A \leftrightarrow B$ and $p_n
\leftrightarrow \bar{p}_n$:
\begin{equation}
  \widetilde{M}_{AB}^{\mathrm{op}}\left(\frac{u_1}{u_2}\Big|q,t\Big| p,\bar p\right) =
  \widetilde{M}_{BA}\left(\frac{u_2}{u_1}\Big|q,t\Big|\bar p,p\right)
  \label{eq:20}
\end{equation}
Since the generalized Macdonald polynomials are actually known explicitly
in many cases \cite{awata2011notes,genMac,FJMM-Bethe,Z}, one can just use (\ref{tildeMvsM}) to evaluate the
first blocks of the ${\cal R}$-matrix, and then promote these
examples to the general formula. This is a much simpler way to get explicit expressions as compared with deducing them
from the universal ${\cal R}$-matrix \cite{FJMM-Bethe,FJMMRmat}, as was suggested
in~\cite{SmiRinst},~\cite{Rmat}, and this will be the
approach we adopt here.

\subsection{Refined topological strings and $\mathcal{RTT}$ relations}
\label{sec:topol-strings-rtt}

Refined topological string theory is a hypothetical string (or, more
probably, M-) theory generalizing the theory of topological
strings. Apart from the string coupling $q = e^{-g_s}$, the refined string
theory depends on an extra deformation parameter $t$, which is related
to the non-self-dual Nekrasov $\Omega$-deformation. In order to reduce it to the
ordinary topological string theory, one should put $t=q$. The
amplitudes of refined strings on the toric Calabi-Yau threefolds have been
computed with the help of the refined topological vertex
technique~\cite{IKV,AK}. The main idea of this technique \cite{3dvertex} is to break
down the threefold into $\mathbb{C}^3$ patches and find the universal
amplitudes, trivalent \emph{refined vertices} on those patches.
Each vertex depends on boundary conditions on three Lagrangian branes
of topology $S^1\times D^2$ sitting on the legs of the toric
diagram. These boundary conditions are encoded in the Young diagram,
which summarizes the winding numbers of string boundaries on the
branes. The final answer for any amplitude, either closed string,
i.e.\ without any branes, or open with nonzero boundary
conditions, is obtained as a sum of the product of topological
vertices over intermediate Young diagrams with ``a propagator''
containing K\"ahler parameters of the edges.

We employ an algebraic approach to the refined topological vertices
developed in~\cite{AFS}. The vertices are treated as intertwiners of
the Fock representations of the DIM algebra, each
representation corresponding to the leg connected to the vertex. The
slopes of the legs are encoded in the central charges of the
corresponding representations. Finally, the sum over intermediate
Young diagram residing on the leg is interpreted as a sum over the
complete basis of states in the corresponding Fock representation. Thus,
to any toric diagram, one associates an intertwiner between tensor
products of Fock representations. Such intertwiners by definition
commute with the action of the DIM algebra on the representations. To
get the answer for the amplitude from the intertwiner, one should
simply evaluate the matrix element of the intertwiner between the
basis vectors in the Fock modules corresponding to the external Young
diagrams (see details and examples in \cite{DIM1}).

The sum over intermediate Young diagrams in the computation of any
amplitude can also be interpreted as a ``network''-type matrix
model~\cite{MMZ,MMZWI,DIM1}. For certain ``balanced'' toric diagrams, the
corresponding matrix model can be identified with the Dotsenko-Fateev (DF)
representation for the multipoint conformal blocks of the $q$-deformed
$W_N$ algebra \cite{DIM1}. Moreover, one can usually obtain \emph{two} such
descriptions related by the action of the spectral duality: either
as a $(k+2)$-point $W_N$-block or as an $(N+2)$-point $W_k$-block, the
corresponding toric diagrams being related to each other by
$\frac{\pi}{2}$ rotation. The existence of two coinciding conformal
blocks of different kinds is related to the AGT duality as shown
in~\cite{Z}.

The fact that any toric diagram essentially represents a contraction
of the intertwiners commuting with the action of the DIM algebra
leads to important implications for matrix model, to the Ward
identities \cite{MMZWI,DIM1}. These identities are very similar to the $W_N$-algebra
Ward identities derived in the DF representations, where the generators of
algebra also commute with the set of \emph{screening charges}
$Q_a$. These identities relate the correlators involving descendants to
those of the primary fields. In fact, one can show that this
construction can be entirely incorporated in the DIM approach to
topological strings. The $W_N$ generators are obtained from the DIM
generators acting on the tensor product of $M$ Fock modules, and the
screening charges arise from a certain combination of the DIM
intertwiners. In the context of gauge theory, such identities were
described in~\cite{qq} as following from the regularity of
$qq$-characters. Also, in the Nekrasov-Shatashvili limit these
identities turn out to give the Baxter TQ equations for the Seiberg-Witten
integrable systems related to the gauge theory \cite{AGTint}.

However, we would like to describe a different form of integrable
structure, related not to \emph{infinitesimal} transformations
(realized as the action of the DIM algebra), but to the ``large'' action
of an automorphism group. This ``large'' action is performed by the
$\mathcal{R}$-matrix which we have describe above. Indeed, the
$\mathcal{R}$-matrix permutes the representations and thus acts on the
intertwiners, i.e.\ on the topological vertices. The refined topological
string amplitudes can then be interpreted as matrix elements of the transfer (or Lax)
matrices, which are permuted according to the $\mathcal{RTT}$-relations. More concretely,
the simplest $\mathcal{T}$-operator taking part in the relations is given by the
following conifold geometry:
\begin{equation}
  \label{eq:88}
    \mathcal{T}_{AB}^{RP} (Q,u,z)
  = \parbox{2.5cm}{\includegraphics[width=2.5cm]{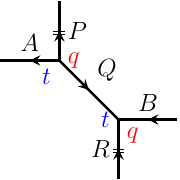}}
\end{equation}

\begin{figure}[t]
  \centering
  \includegraphics[width=12cm]{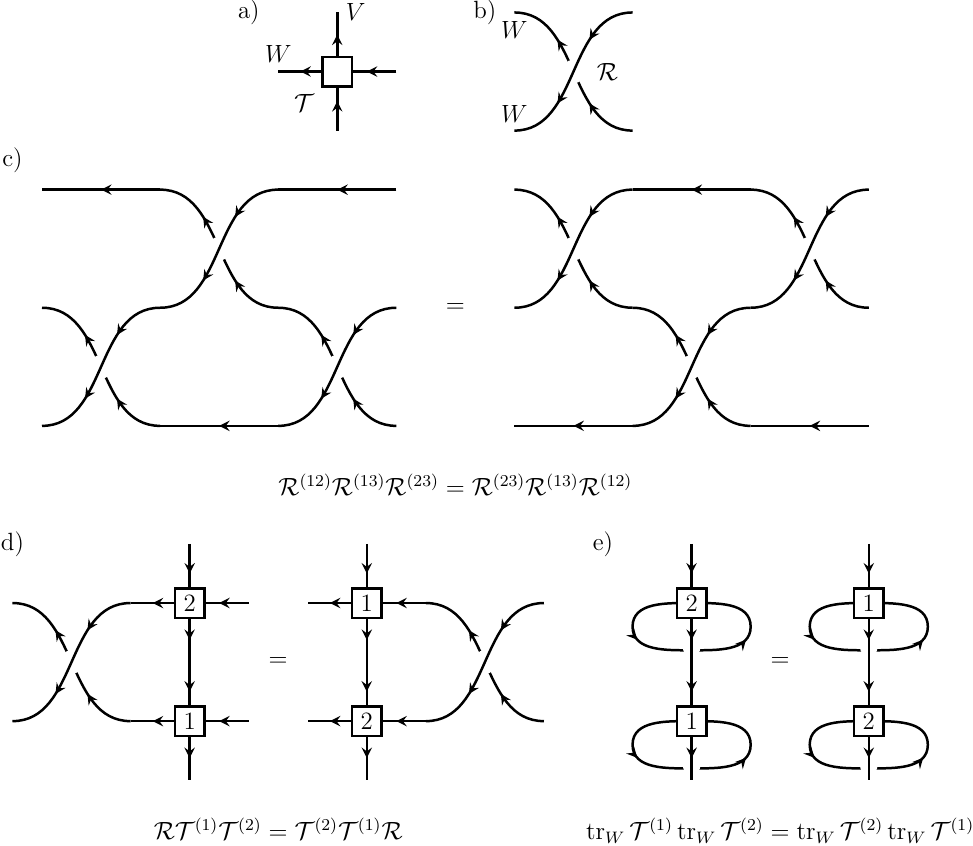}
  \caption{Commuting integrals of motion in a quantum integrable system can be constructed using two essential
    building blocks: a) the $\mathcal{T}$-operator acting in the
    tensor product of the \emph{quantum} space $V$ (vertical leg) and the \emph{auxiliary} space $W$ (horizontal leg), b)
    the $\mathcal{R}$-matrix acting on $W \otimes W$ and satisfying c),
    the Yang Baxter equation. d) $\mathcal{R}$ and $\mathcal{T}$ have
    to satisfy the $\mathcal{RTT}$ relations, providing
    commutation relations for the $\mathcal{T}$-operators. e) Taking
    the trace of the $\mathcal{T}$-operator over the auxiliary space,
    one gets commuting operators acting in the quantum space, these are
    the quantum integrals of motion.}
  \label{fig:2}
\end{figure}

The action of the $\mathcal{R}$-matrix on the toric diagram
$\mathcal{T}$-operator is given by Eq.~\eqref{eq:15}. The whole toric
diagram now looks like the combination of objects familiar from the
theory of quantum integrable models (e.g.\ spin chains): the
$\mathcal{R}$-matrices and $\mathcal{T}$-operators (see
Fig.~\ref{fig:2}). The vertical representations are identified with the
\emph{quantum} spaces (e.g.\ Hilbert spaces of the spins),
while the horizontal ones are the \emph{auxiliary} spaces, on which
the $\mathcal{R}$-matrix acts. In terms of quantum group elements, the quantum space is associated with the algebra of functions, while the auxiliary one with the universal enveloping algebra \cite{qge}.
Geometrically the $\mathcal{R}$-matrix
performs a generalized version of the flop transition on the Calabi-Yau
manifold \cite{flop}.

Let us also make a remark on a relation between the spectral duality
and the $\mathcal{R}$-matrix. The spectral duality $\mathcal{S}$
rotates the lattice of generators (or the preferred direction on the
toric diagram) in Fig.~\ref{fig:1} by $\frac{\pi}{2}$. It turns out
that the $\mathcal{R}$-matrix can be naturally interpreted using the
$\mathcal{S}$ automorphism. As we have already seen, the
$\mathcal{R}$-matrix looks simple in the basis of generalized
Macdonald polynomials: indeed, it is just the permutation of the
spaces and the spectral parameters denoted by the $\mathrm{op}$ label
in~(\ref{eq:20}). The generalized Macdonald basis is spectral dual to that
of tensor products of the ordinary Macdonald polynomials. Thus, to compute the
$\mathcal{R}$-matrix in the basis of ordinary polynomials, one should first
rotate to the spectral dual frame using $\mathcal{S}$, then make the
permutation of the spaces and finally rotate back using
$\mathcal{S}^{-1}$. We thus obtain the relation of the form
\begin{equation}
  \label{eq:22}
  \mathcal{R} = \mathcal{S}^{-1} \sigma \mathcal{S}
\end{equation}
where $\sigma$ denotes the permutation of representations or legs
of the toric diagram. We will encounter this relation when performing
concrete computations of the $\mathcal{R}$-matrices.

Having transfer matrices, one can take traces of them. Just
as in any quantum integrable system, these traces generate a family
of commuting integrals of motion. Those too have an interpretation in
terms of topological string. However, this time one has to
\emph{compactify} the toric diagram, i.e.\ to consider not the toric Calabi-Yau
threefold, but its compactified version. The situation here
resembles that considered in the classic paper by V.Bazhanov, S.Lukyanov
and A.Zamolodchikov~\cite{BLZ}, where an infinite family of integrals
of motion in CFT was derived. The intertwiners of DIM play the
role of exponentials of free fields and their traces, i.e.
compactifications provide the integrals of motion. In the language of matrix
models, this corresponds to further deforming the measure:
depending on the direction of compactification, it becomes either
\emph{elliptic} or \emph{affine.}  Eventually, the matrix elements of
commuting integrals of our integrable system correspond to certain
correlators in the elliptic or affine matrix models. There are different
directions to pursue from this point. However, we only sketch possible
further developments in sec.~\ref{sec:conclusion}.

Let us point out an important difference between our approach and
several recent works dealing with the DIM Ward identities and
$\mathcal{R}$-matrices~\cite{Matsuo,FJMM-Bethe,FJMMRmat,Rmat}. We will predominantly work
with \emph{horizontal} representations, whereas in~\cite{Matsuo,FJMM-Bethe,FJMMRmat,Rmat}
it was essential to consider the \emph{vertical} representations. It
would be very interesting to unify the two approaches and make the
$SL(2,\mathbb{Z})$ invariance and duality between vertical and horizontal directions manifest.

We understand that similar calculations for DIM $\mathcal{R}$-matrix have also been done by
S.Shakirov~\cite{ShamRmat}.

\section{$\mathcal{R}$-matrices: from $\beta$-deformation to
  $(q,t)$-deformation}
\label{sec:mathc-matr-diff}

In this section we implement the algorithm given in
sec.~\ref{sec:dim-algebra-gener} to compute the DIM
$\mathcal{R}$-matrix.

To warm up, we start with two simplified examples. The first one
(sec.~\ref{sec:triv-exampl-t=q}) is the trivial case of
\emph{unrefined} topological string, i.e.\ $t=q$. The second one is
the ``$4d$ limit'' of the DIM algebra, the affine Yangian
$Y(\hat{\mathfrak{gl}}_1)$ considered in
sec.~\ref{sec:affine-yang-gener} (see \cite{Yangian,SV,MaOk,SmiRinst,Smirnov2,M1,M2,Matsuo}). The construction in this case is
parallel to the DIM algebra, with the generalized Macdonald polynomials
replaced by the generalized Jack polynomials, and this makes the
formulas a bit less bulky. Finally, in
sec.~\ref{sec:dim-mathcalr-matrix} we turn to our real focus, the
DIM $\mathcal{R}$-matrix.

\subsection{A trivial example: $t=q$, Schur polynomials}
\label{sec:triv-exampl-t=q}
For the unrefined topological string, i.e.\ for $t=q$ the generalized
Macdonald Hamiltonian $\hat{\mathcal{H}}_1$,~(\ref{Ham}) degenerates
into the sum of two noninteracting Ruijsenaars Hamiltonians. Thus, the
generalized Macdonald polynomials become just the product of two Schur
functions, and do not essentially depend on the spectral parameters
$u_{1,2}$. This means that Eq.~(\ref{tildeMvsM}) defines a
\emph{trivial} $\mathcal{R}$-matrix, which is proportional to the
identity matrix\footnote{There are different conventions on numbering
  the strands \emph{entering} and \emph{exiting} from the
  $\mathcal{R}$-matrix. In the theory of integrable systems, it is
  standard to label strands according to their spectral
  parameters. However, in knot theory, one usually assigns numbers to
  the \emph{positions} of strands in the slice. We use the first
  choice, and the second one can be obtained by taking a product of
  $\mathcal{R}$ and the matrix of permutation of two strands
  $\sigma_{12}$.}:
\begin{equation}
  \label{eq:19}
  \mathcal{R}{}_{AB}^{CD}(u)|_{t=q} \propto \delta_A^C \delta_B^D.
\end{equation}

\subsection{Affine Yangian $\mathcal{R}$-matrix from generalized Jack
  polynomials}
\label{sec:affine-yang-gener}

\paragraph{Two strands.}
\label{sec:two-strands}
For the tensor product of two Fock representation, the generalized Jack
polynomials are eigenfunctions of the $\beta$-deformed cut-and-join
operator $\hat {\cal H}^{(\beta)}_1$ which belongs to the Cartan
subalgebra of the affine Yangian:
\begin{multline}
\mathcal{H}^{(\beta)} =
\frac{1}{2}
\sum_{n,m=1}^\infty \left(\beta(n+m)p_np_m\frac{\p}{\p p_{n+m}}+nmp_{n+m}\frac{\p^2}{\p p_n\p p_m}\right)
+ \frac{1}{2}\sum_{n=1}^\infty \Big(2u+(\beta-1)(n-1)\Big) np_n\frac{\p}{\p p_n} + \\
+ \frac{1}{2}
\sum_{n,m=1}^\infty \left(\beta(n+m)\bar p_n\bar p_m\frac{\p}{\p \bar p_{n+m}}+
nm\bar p_{n+m}\frac{\p^2}{\p \bar p_n\p \bar p_m}\right)
+ \frac{1}{2}\sum_{n=1}^\infty \Big(2\bar u+(\beta-1)(n-1)\Big) n\bar p_n\frac{\p}{\p \bar p_n} + \\
+ (1-\beta)\sum_{n=1}^\infty n^2\bar p_n\frac{\p}{\p p_n}\label{eq:67}
\end{multline}
It is the last term in the third line which breaks the symmetry
between $p$ and $\bar p$ and makes dual polynomials different.  Notice
that this term vanishes for $\beta=1$, i.e.\ in the trivial case that we have
considered in the previous subsection. In general, the eigenvalues corresponding to the
eigenfunctions $J_{AB}\{p,\bar p\}$
are
\begin{equation}
  \kappa_{AB}^{(\beta)} = \sum_{(i,j)\in A} \Big(u +
  (i-1)-(j-1)\beta\Big) + \sum_{(i,j)\in B} \Big(\bar u +
  (i-1)-(j-1)\beta\Big)\label{eq:53}
\end{equation}
Notice also that the eigenvalues~(\ref{eq:53}) are degenerate:
e.g.\
$\kappa^{(\beta)}_{[1,1],[2]}=\kappa^{(\beta)}_{[2],[1,1]}$. Thus,
one still needs higher Hamiltonians $\mathfrak{\mathcal{H}}_n^{\beta}$ with $n \geq 2$
to uniquely specify the polynomials, which
makes the problem a little sophisticated.  As we have already
mentioned, this is cured at the DIM level, where the $(q,t)$-deformation
makes the eigenvalues non-degenerate.

One can take the answer for the eigenfunctions from \cite{MoSmi}. The first
level reads:
\begin{gather}
  J_{[1],\emptyset} = (1-\beta) \bar p_1-(\bar{u}-u) p_1, \qquad
   J^{*}_{[1],\emptyset} = (1+u-\bar{u}-\beta) p_1 \notag\\
  J_{\emptyset,[1]} = (u-\bar{u}-1+\beta)\bar p_1, \qquad J^{*}_{\emptyset,[1]} =
  (u-\bar{u})\bar p_1 + (1-\beta) p_1.
\end{gather}

It is now straightforward to obtain the ${\cal R}$-matrix from the
relation similar to Eq.~(\ref{tildeMvsM}). However, first, we emphasize a subtlety which makes the definition of the
$\mathcal{R}$-matrix nontrivial. The point is the simplicity of
the definition of the ``opposite'' polynomials~(\ref{eq:20}). This
definition in fact depends on the choice of the particular
\emph{special normalization} of the polynomials. To put it another way,
the $\mathcal{R}$-matrix indeed transforms each generalized Jack
polynomial into the corresponding ``opposite'' polynomial, however,
the coefficient needs not necessarily to be the identity. Thus, for arbitrary
normalization of the generalized polynomials, one has the following
definition of the opposite ones:
\begin{equation}
  \label{eq:21}
  \frac{N_{AB}(u_1-u_2|\beta)}{N_{BA}(u_2 - u_1|\beta)}    J_{AB}^{\mathrm{op}}\left(u_1-u_2|\beta| p,\bar p\right) =
  J_{BA}\left(u_2-u_1|\beta|\bar p,p\right),
\end{equation}
where $N_{AB}(u|\beta)$ is the normalization coefficient
absent for the special normalization. Then, the $\mathcal{R}$-matrix is
indeed given by
\begin{equation}
  \label{eq:23}
      J_{AB}^{\mathrm{op}}\left(u_1-u_2| \beta | p,\bar p\right) =
  \sum_{C,D} {\cal R}^{CD}_{AB}\!\left(u_1-u_2\right)\cdot
  J_{CD}\left(u_1-u_2|\beta | p, \bar p\right)
\end{equation}
or, using the Jack scalar product,
\begin{equation}
  \label{eq:24}
  {\cal R}^{CD}_{AB}\!\left(u_1-u_2\right) = \frac{1}{||J_{AB}^{\mathrm{op}}(u_1 - u_2)||^2}\langle J^{*}{}_{AB}^{\mathrm{op}}\left(u_1-u_2| \beta | p,\bar p\right)|
  J_{CD}\left(u_1-u_2|\beta | p, \bar p\right) \rangle.
\end{equation}
The Jack scalar product is defined as
\begin{equation}
  \label{eq:47}
  \langle f(p_n) | g(p_n) \rangle = f \left( \frac{n}{\beta} \frac{\partial}{\partial p_n} \right) g(p_n) |_{p_n=0}
\end{equation}
Notice the conjugate polynomial $J^{*}{}_{AB}^{\mathrm{op}}$ in the bra vector in
Eq.~(\ref{eq:24}).

We now describe the special normalization of Jack polynomials
explicitly. To this end, we expand $J_{AB}$ in the basis of monomial
symmetric functions:
\begin{equation}
  \label{eq:48}
  J_{AB}(u|\beta|p_n,\bar{p}_n) = N_{AB}(u|\beta) m_A(p_n) m_B(\bar{p}_n)
  + \sum_{CD\neq AB} C_{AB}^{CD}(u|\beta) m_C(p_n) m_D(\bar{p}_n)
\end{equation}
where $m_A(p_n)$ denote the monomial symmetric polynomials and the
normalization factor is
\begin{equation}
  \label{eq:51}
  N_{AB}(u|\beta) = g_{AB}(u|\beta) \prod_{(i,j)\in
    A}\left( A_i - j + \beta (A^{\mathrm{T}}_i - j + 1) \right)\prod_{(i,j)\in
    B}\left( B_i - j + \beta (B^{\mathrm{T}}_i - j + 1) \right)
\end{equation}
and
\begin{equation}
 g_{AB} (x)= \prod_{(i, j) \in A} \left(  x + A_i - j + \beta
   (B^{\mathrm{T}}_j - i + 1) \right) \prod_{(i,j) \in B}\left( x - B_i + j - 1 - \beta(A^{\mathrm{T}}_j - i) \right)
\end{equation}
is the usual $4d$ Nekrasov factor. The
normalization factor $N_{AB}$ in Eq.~(\ref{eq:48}) is the same as in
Eq.~(\ref{eq:21}). Notice that the \emph{special} normalization is
different from another popular choice of normalization, which we call
\emph{standard}. In the standard normalization, the coefficient in front of
$m_A(p_n) m_B(\bar{p}_n)$ in $J_{AB}$ is unit and $||J_{AB}||^2 =
||J_A||^2 ||J_B||^2$ is independent of $u$.

The normalization factors satisfy the
identity $N_{AB}(u|\beta) N_{BA}(-u|\beta) ||J_A||^2 ||J_B||^2 =
z^{\mathrm{vec}}_{AB}(u|\beta)$ where $z^{\mathrm{vec}}$ is the vector
contribution to the Nekrasov functions \cite{Nek,NekOk}. In particular, one has
\begin{equation}
  \label{eq:52}
  ||J_{AB}||^2 = z^{\mathrm{vec}}_{AB}(u|\beta).
\end{equation}
The polynomials~(\ref{eq:21}) are already written in the special
normalization. This normalization is in fact natural from the
cohomological point of view. The generalized Jack polynomials can be
associated to the fixed points in the moduli space of $SU(2)$
instantons (or to the Hilbert schemes of points on $\mathbb{C}^2$) \cite{Smirnov2}. The
action of the first Hamiltonian $\hat{\mathcal{H}}_1$ is given by the
cup product with the first Chern class in the cohomology, whereas
higher Hamiltonians are cup products with higher Chern classes. They
commute simply because of the commutativity of the cup product. The
specially normalized generalized Jack polynomials then describe stable
envelopes of the corresponding fixed points.

Having understood the subtle point of normalization, we get the
$\mathcal{R}$-matrix in the basis of generalized Jack polynomials:
\begin{equation}
  {\cal R}^{(\beta)} = \left(\begin{array}{cc} 1-\eta &
      \frac{\eta}{\eta + 1}
      \\ \eta - \eta^2 & \frac{\eta^2 + 1}{\eta+1}
    \end{array}\right)
  \label{2x2Rbeta}
\end{equation}
with
\begin{equation}
  \eta = \frac{1-\beta}{u-\bar u}
\end{equation}
This $\mathcal{R}$-matrix, though simple, is still nontrivial. It
should be supplemented with the identity block arising from the
generalized Jack polynomials at the zeroth level:
\begin{equation}
  \label{eq:54}
  J_{\varnothing,\varnothing}(u|\beta|p, \bar{p}) =
  J^{*}{}_{\varnothing,\varnothing}(u|\beta|p, \bar{p}) = 1
\end{equation}
The resulting $3 \times 3$ matrix
\begin{equation}
  \label{eq:55}
  \left(\begin{array}{ccc}1 & 0 & 0\\
      0 &1-\eta &
      \frac{\eta}{\eta + 1}
      \\ 0& \eta - \eta^2 & \frac{\eta^2 + 1}{\eta+1}
    \end{array}\right)
\end{equation}
should satisfy some form of the Yang-Baxter relation. However, it
does not look like the $3 \times 3$ block of the standard rational
$\mathcal{R}$-matrix
\begin{equation}
  \label{eq:56}
  \left(\begin{array}{ccc}1 & 0 & 0\\
      0 &\frac{1}{1+\eta} &
      \frac{\eta}{1+\eta}
      \\ 0& \frac{\eta}{1+\eta} & \frac{1}{1+\eta}
    \end{array}\right)
\end{equation}
which is the only $3 \times 3$ rational solution to the Yang-Baxter
equation. This discrepancy is resolved if we recall that the basis of generalized
Jack polynomials in the tensor product of two Fock
representations of the affine Yangian, are not factorised into vectors
in each representation. To get more familiar expression for the
$\mathcal{R}$-matrix, we should consider its matrix elements in a
basis, where the vectors are factorized into tensor products, e.g.\ the
products of the Jack polynomials $J_A^{(\beta)}(p_n)
J_B^{(\beta)}(\bar{p}_n)$. The basis is changed with the help of the
generalized Kostka matrices:
\begin{equation}
  \label{eq:57}
  K_{AB}^{CD}(u|\beta) = \langle
  J^{*}{}_{AB}(u|\beta)| (|J_C^{(\beta)} \rangle \otimes
  |J_D^{(\beta)} \rangle), \qquad   K^{*}{}_{AB}^{CD}(u|\beta) =
  (\langle J_A^{(\beta)} | \otimes
  \langle J_B^{(\beta)}|)| J_{CD}(u|\beta)\rangle
\end{equation}
At the first level, we have
\begin{equation}
  \label{eq:59}
  K(u|\beta)  =\left(
    \begin{array}{cc}
      1 & 0\\
      \eta & 1
    \end{array}
  \right), \qquad  K^{*}(u|\beta) =\left(
    \begin{array}{cc}
      1 & 0\\
      -\eta & 1
    \end{array}\right)
\end{equation}
The $\mathcal{R}$-matrix in the factorized basis of the ordinary Jack
polynomials is given by
\begin{equation}
  \label{eq:58}
  \mathcal{R}^{(\beta)}_{\text{ord Jack}} = K^{*} \frac{1}{||J||^2}
  \mathcal{R}^{(\beta)} \frac{1}{||J||^2} K = \frac{1}{\eta + 1}\left(
    \begin{array}{cc}
      1 & \eta\\
      \eta & 1
    \end{array}
  \right),
\end{equation}
where $||J||^2$ denotes the diagonal matrix containing the norms of
generalized Jack polynomials. Eq.~(\ref{eq:58}) gives the standard
rational $\mathcal{R}$-matrix~(\ref{eq:56}). Formula~(\ref{eq:58})
can be understood as a decomposition of the $\mathcal{R}$-matrix
into the upper and lower triangular parts, since
\begin{equation}
  \label{eq:60}
  K^{*} \frac{1}{||J||^2} \mathcal{R}^{(\beta)} = \left(
    \begin{array}{cc}
      1-\eta &  \frac{\eta}{\eta + 1}\\
      0 & \frac{1}{\eta + 1}
    \end{array}
\right) \quad \text{and} \quad K = \left(
    \begin{array}{cc}
      1 & 0\\
      \eta & 1
    \end{array}
  \right)
\end{equation}
Moreover, one can refine this decomposition even further: one can
identify the upper and lower triangular parts \emph{with identities on
  the diagonal} and the diagonal part sandwiched between them. To
obtain this decomposition, one should explicitly write down the
normalization coefficients $N_{AB}(u|\beta)$ of generalized Jack
polynomials in the formulas for the $\mathcal{R}$-matrix. The diagonal
part then comes from the term $\frac{N_{AB}(u_1-u_2|\beta)}{N_{BA}(u_2
  - u_1|\beta)}$ in Eq.~(\ref{eq:21}) and the whole expression
becomes
\begin{equation}
  \label{eq:61}
  \mathcal{R}^{(\beta)}_{\text{ord Jack}} = \left\{ K^{*} \frac{1}{||J||^2}
  \mathcal{R}^{(\beta)} \mathcal{N}^{-1} \right\}\cdot  \mathcal{N} \cdot K
\end{equation}
where $\mathcal{N}_{AB}^{CD} = \frac{N_{AB}(u_1-u_2|\beta)}{N_{BA}(u_2
  - u_1|\beta)} \delta_A^C \delta_B^D = \left(
  \begin{smallmatrix}
    1-\eta& 0\\
    0 & -(1+\eta)^{-1}
  \end{smallmatrix}
\right)$. In Eq.~(\ref{eq:61}) the term in the curly brackets is
upper triangular with identities on the diagonal, the matrix
$\mathcal{N}$ is diagonal and $K$ is lower triangular.

Decomposition of the $\mathcal{R}$-matrix into the upper and lower
triangular parts also has an interpretation in the cohomology of the
instanton moduli space~\cite{SmiRinst,Rmat}. Parts of the
$\mathcal{R}$-matrix decomposition correspond to \emph{stable
  envelopes} of the fixed points, i.e.\ to the cohomology classes of the
attracting domains of the fixed points under a certain
$\mathbb{C}^{\times}$-action. The $\mathcal{R}$-matrix in this
approach is given by the infinite product of the ``wall
$\mathcal{R}$-matrices'' labelled by rational slopes (determined by
the integer pairs corresponding to the double gradings of DIM or affine Yangian generators, as
depicted in Fig.~\ref{fig:1}) within the interval of angles: $[0,\pi]$. The
lower triangular part corresponds to the product over
$[0,\frac{\pi}{2}]$, the diagonal one represents the wall with infinite
slope, and the lower triangular matrix is the product over
$[\frac{\pi}{2},\pi]$:
\begin{equation}
  \label{eq:62}
  \mathcal{R}^{(\beta)} = \mathcal{R}_{[\frac{\pi}{2},\pi]}^{(\beta)}
  \mathcal{R}_{\infty}^{(\beta)} \mathcal{R}_{[0,\frac{\pi}{2}]}^{(\beta)}.
\end{equation}

Such a decomposition is just a reflection of the identity $\mathcal{R}
= \mathcal{S} \sigma \mathcal{S}$ which we have mentioned in the
Introduction. Each wall $\mathcal{R}$-matrix corresponds to a change
of the preferred direction from one ``chamber'' in
$(\mathbb{C}^{\times})^2$ to another, the border between them being
the line of rational slope. The product of wall
$\mathcal{R}$-matrices over angles $[0,\frac{\pi}{2}]$ is nothing but
the automorphism $\mathcal{S}$. Also, in~\cite{MZDecompNekr} it was
shown that the generalized Kostka matrices are in fact the matrix elements
of $\mathcal{S}$ in the basis of eigenfunctions of the DIM Cartan
subalgebra (the story for the affine Yangian, which we study in this
section, is parallel). Depending on whether the preferred direction
(or the representation in question) is horizontal or vertical, the
eigenfunctions of the Cartan subalgebra can be either ordinary or generalized
polynomials. $\mathcal{S}$ performs a linear transformation between
the two basis sets, and is thus nothing but the Kostka matrix $K$ as
clearly seen from the definition~(\ref{eq:57}). Eventually, the
decomposition~(\ref{eq:61}) is a reflection of the
decomposition~(\ref{eq:22}), where $\sigma$ is accompanied by
multiplication with the diagonal matrix $\mathcal{N}$.

Yet another meaning of the $\mathcal{R}$-matrix that we have just obtained
can be seen by noticing that the affine Yangian acting on the tensor
product of two Fock modules contains the Virasoro subalgebra generated
by the dressed current $t(z) = \alpha(z) x^{+}(z) \beta(z)$. A pair
of Heisenberg algebras provides a bosonization of this Virasoro
algebra. However, it is well-known that there are two such
bosonizations related to each other by the Liouville reflection
matrix \cite{ZZ}. The job of the reflection matrix is similar to that of the
$\mathcal{R}$-matrix: it exchanges the two types of bosons. Indeed,
one can see that the two objects are in fact one and the same. For
example, as we have discussed in the previous subsection, the
$\mathcal{R}$-matrix is trivial for $t=q$ or equivalently $\beta=1$,
and the Liouville reflection matrix is also trivial, since $c = 1 + 6
(\sqrt{\beta} - 1/\sqrt{\beta})^2 = 1$ in this case. In fact, the
tensor product of two Heisenberg algebras acting on two Fock modules
contains in addition to the Virasoro also the diagonal Heisenberg
subalgebra, which is usually called the ``$U(1)$ part'' in the AGT
context \cite{CarlssonOk}. This part is of course left invariant by the reflection
matrix. One can see that the $\mathcal{R}$-matrix also leaves this
subspace invariant. Thus, the $\mathcal{R}$-matrix of the affine
Yangian is nothing but the reflection matrix of the Liouville theory.

\paragraph{More strands.}
\label{sec:many-strands}
For more than two strands, the generalized Jack polynomials can still be
described as eigenfunctions of a certain Hamiltonian
$\mathcal{H}_1^{(\beta)}$ sitting inside the affine
Yangian,~\cite{MoSmi}. The polynomials $J_{\vec A}\{\vec p_n\}$
in this case depend on $r$ sets of time-variables $p^{(k)}_n$,
$k=1,\ldots,r$, on $r$ Young diagrams $\vec A = \{A_1,\ldots,A_r\}$
and on $r$ spectral parameters $\vec u = \{u_1,\ldots,u_r\}$.  The
Hamiltonian is now a linear combination
\begin{equation}
  {\cal H}^{(\beta)}_1 = \sum_{k=1}^r {\cal
    H}^{(\beta)}_{(k)} + (1-\beta)\sum_{k_1<k_2} {\cal
    H}^{(\beta)}_{(k_1,k_2)}
\end{equation}
with
\begin{equation}
\!{\cal H}^{(\beta)}_{(k)} = \frac{1}{2}
\sum_{n,m=1}^\infty \!\!\left(\!\beta(n+m)p_n^{(k)}p_m^{(k)}\frac{\p}{\p p_{n+m}^{(k)}}+
nmp_{n+m}^{(k)}\frac{\p^2}{\p p_n^{(k)}\p p_m^{(k)}}\right)
\!+ \frac{1}{2}\sum_{n=1}^\infty \Big(2u_k+(\beta-1)(n-1)\Big) np_n^{(k)}\frac{\p}{\p p_n^{(k)}}
\end{equation}
and
\begin{equation}
{\cal H}^{(\beta)}_{(k_1,k_2)} = \sum_{n=1}^\infty n^2p_n^{(k_1)}\frac{\p}{\p p_n^{(k_2)}}
\end{equation}
The construction of the $\mathcal{R}$-matrix is similar to the case of
two strands. The important difference is that there are now $r-1$
$\mathcal{R}$-matrices, which permute the factors in the tensor
product of Fock modules. They form a representation of the $r$-strand
braid group ${\cal B}_r$.

In the basis of generalized Jack polynomials, the resulting
$\mathcal{R}$-matrices look rather ugly (see
Appendices~\ref{sec:beta-deform-vers} and
especially~\ref{sec:beta-deform-vers-1}). However, in the basis of
ordinary Jack polynomials, the expressions simplify. In this basis, the
$\mathcal{R}$-matrix acting on each pair of strands becomes a copy of
the two-strand $\mathcal{R}$-matrix:
\begin{equation}
  \label{eq:74}
  \mathcal{R}^{(\beta)}_{ij} =\sum_{A, B, C, D} \mathcal{R}^{(\beta)}_{\text{ord Jack}}{}_{AB}^{CD}(u_i-u_j)\,\,
  \mathrm{id} \otimes \cdots \otimes \mathrm{id} \otimes |J_C\rangle_i \langle J_A|_i \otimes
  \cdots \otimes |J_C\rangle_j \langle J_A|_j \otimes \mathrm{id} \otimes \cdots \otimes \mathrm{id}
\end{equation}
Thus, all the familiar results from integrable systems hold, e.g.\
the fusion of $\mathcal{R}$-matrices.  For three strands, one can also
check the Yang-Baxter equation and it works as expected. The relation
with the spectral duality~(\ref{eq:22}) for several strands is modified in
an obvious way:
\begin{equation}
  \label{eq:63}
  \mathcal{R}^{(\beta)}_{i,j} = \mathcal{S}^{-1} \sigma_{i,j} \mathcal{S},
\end{equation}
where $\sigma_{i,j}$ permutes the $i$-th and $j$-th strands.

\subsection{DIM $\mathcal{R}$-matrix from generalized Macdonald
  polynomials}
\label{sec:dim-mathcalr-matrix}
In this section, we compute the DIM $\mathcal{R}$-matrix from the generalized
Macdonald polynomials. This turns out to be simpler and more natural
than the affine Yangian $\mathcal{R}$-matrix in the previous subsection.

\paragraph{Two strands.}
\label{sec:two-strands-1}
As we described in the Introduction,
the generalized Macdonald polynomials are eigenfunctions of the element
$x^{+}_0$ of the DIM algebra acting in the tensor product of Fock
representations:
\begin{equation}
  \label{eq:66}
  \mathcal{H}_1 \widetilde{M}_{AB} = \kappa_{AB}\widetilde{M}_{AB}
\end{equation}
where
\begin{equation}
  \label{eq:65}
    \kappa_{AB} = u_1 \sum_{i\geq 1} q^{A_i} t^{-i} + u_2 \sum_{i\geq 1}
  q^{B_i} t^{-i}
\end{equation}
and
\begin{multline}
  \label{eq:11}
{\cal H}_1 =  \rho_{u_1} \otimes \rho_{u_2} \Delta (x^{+}(z))
= \oint \frac{dz}{z} \left[ u_1 \Lambda_1 (z) + u_2 \Lambda_2 (z) \right] =\\
  =\oint \frac{dz}{z} \Biggl[ u_1 \exp \left( \sum_{n \geq 1}
    \frac{1-t^{-n}}{n} p^{(1)}_n z^{-n} \right) \exp \left( \sum_{n
      \geq 1} \frac{1-q^n}{n} z^n \frac{\partial}{\partial p^{(1)}_n}
  \right)+\\
  + u_2 \exp \left( \sum_{n \geq 1} \frac{1-t^{-n}}{n} \left( (1 -
      t^n/q^n) p^{(1)}_n + p^{(2)}_n(q/t)^{-n/2} \right) z^{-n}
  \right) \exp \left( \sum_{n \geq 1} \frac{1-q^n}{n} z^n (q/t)^{n/2}
    \frac{\partial}{\partial p^{(2)}_n} \right) \Biggr]
\end{multline}
where $\Delta$ is the DIM coproduct and $\rho_u$ denotes the horizontal
Fock representation \cite{DIM,DIM1}. Note that the eigenvalues~(\ref{eq:65}) are
\emph{non-degenerate} and, though there are higher Hamiltonians
$\mathcal{H}_n$ (see Appendix B), they are not needed to determine the
spectrum. One can ask how does the degeneration appear in the Yangian
limit $q \to 1$. The Hamiltonian $\mathcal{H}_1$ in this limit is
expanded in series of operators in $(q-1)$, and the first term is the
Hamiltonian $\mathcal{H}^{(\beta)}_1$~(\ref{eq:67}) which we considered in
the previous subsection. Since this is just the first term in the
expansion, some eigenvalues degenerate and one needs higher
Hamiltonians $\mathcal{H}_n^{(\beta)}$. However, all these
Hamiltonians are contained in the expansion of $\mathcal{H}_1$.

We slightly changed our notations compared to the previous
subsection: the order of Young diagrams $A$, $B$ is reversed as compared
to sec.~\ref{sec:affine-yang-gener}. This is done mostly to conform
with the existing literature on the subject, where the discrepancy
seems to be already entrenched.

The generalized Macdonald polynomials at the first level are given
by\footnote{We again remind the reader of the change of convention $Q
  \to Q^{-1}$ as compared to~\cite{Z}.}
\begin{gather}
  \widetilde{M}_{[1], [] } = (1-t) \left(1- \frac{t}{q}
    Q\right)p_1,\qquad
  \widetilde{M}^{*}_{[1], [] } = (1-q)(1-Q) p_1 - (1-q)\left(1-\frac{t}{q}\right) \bar{p}_1\notag\\
  \widetilde{M}_{[],[1]} = (1-t)(1-Q) \bar{p}_1 +
  (1-t)\left(1-\frac{t}{q}\right)p_1, \qquad
  \widetilde{M}^{*}_{[],[1]} = (1-q) \left(1- \frac{t}{q}
    Q\right)\bar{p}_1\notag
\end{gather}
As in the previous subsection, these polynomials are written in the
special normalization such that the definition of opposite polynomials
is given by Eq.~\eqref{eq:20}. In~\cite{Z} a different
normalization was used such that $M_{AB} = 1 \cdot m_A(p_n)
m_B(\bar{p}_n) + \ldots$ We conform with the previous notation and
denote the specially normalized polynomials by $\widetilde{M}_{AB}$
as in~\cite{Z}, Eq.(19). Let us write down the normalization
coefficient, which we take from~\cite{Z}:
\begin{equation}
  \label{eq:73}
  N_{AB}(u|q,t) = G_{BA}(u^{-1}|q,t) C_A (q,t) C_B (q,t)
\end{equation}
where
\be
C_\lambda (q,t) = \prod_{(i,j)\in
    \lambda} \Big(1-q^{\lambda_i-j}t^{\lambda^T_j-i+1}\Big),
    \ee
    \be\label{G}
  G_{AB} (u|q,t)= \prod_{(i, j) \in A} \left( 1 - u q^{A_i - j}
    t^{B^{\mathrm{T}}_j - i + 1} \right) \prod_{(i,j) \in B}\left(1 -
    u q^{-B_i + j - 1} t^{-A^{\mathrm{T}}_j + i} \right) =\\
 = \prod_{(i, j) \in B} \left( 1 - u q^{A_i - j}
    t^{B^{\mathrm{T}}_j - i + 1} \right) \prod_{(i,j) \in A}\left(1 -
    u q^{-B_i + j - 1} t^{-A^{\mathrm{T}}_j + i} \right).\nn
\ee
For the polynomials $M_{AB}$ (\emph{without} tilde), the definition of
opposite polynomial is
\begin{equation}
  \label{eq:81}
  \frac{N_{AB}(u_1/u_2|q,t)}{N_{BA}(u_2/u_1|q,t)} M_{AB}^{\mathrm{op}}\left(u_1/u_2|q,t| p,\bar p\right) =
  M_{BA}\left(u_2/u_1|q,t|\bar p,p\right),
\end{equation}

After the generalized polynomials are found, the $\mathcal{R}$-matrix
is determined in the same way as for the affine Yangian. We simply write
down the main formulas, since the discussion is very similar to
the previous section. The $\mathcal{R}$-matrix in the basis of
generalized Macdonald polynomials is
\begin{equation}
  \label{eq:68}
  {\cal R}^{CD}_{AB}\!\left(\frac{u_1}{u_2}\right) =
  \frac{1}{||\widetilde{M}_{AB}^{\mathrm{op}}\left(\frac{u_1}{u_2}\right)||^2}\left\langle
    \widetilde{M}^{*}{}_{AB}^{\mathrm{op}}\left(\frac{u_1}{u_2}\Big| q,t \Big|
      p,\bar p\right) \Big|
    \widetilde{M}_{CD}\left(\frac{u_1}{u_2}\Big|q,t \Big| p, \bar p\right) \right\rangle
\end{equation}
It is given by
\begin{equation}
  \label{eq:70}
  \mathcal{R} = \left(
\begin{array}{cc}
 -\frac{u_1 \left(u_1^2 q^2+u_2^2 q^2-u_1 u_2 q^2-2 t u_1 u_2 q+t^2 u_1
   u_2\right)}{q \left(u_1-u_2\right) u_2 \left(q u_1-t u_2\right)} &
   \frac{(q-t) u_1 \left(q u_2-t u_1\right)}{q^2 \left(u_1-u_2\right){}^2} \\
 -\frac{(q-t) u_1^2}{u_2 \left(q u_1-t u_2\right)} & \frac{u_1 \left(q u_2-t
   u_1\right)}{q \left(u_1-u_2\right) u_2} \\
\end{array}
\right)
\end{equation}

Transformation to the basis of ordinary Macdonalds is performed using
the $q$-deformed versions of generalized Kostka matrices:
\begin{equation}
  \label{eq:71}
  K = \left(
\begin{array}{cc}
 1 & \frac{(q-t) u_2}{q \left(u_1-u_2\right)} \\
 0 & 1 \\
\end{array}
\right),\qquad K^{*} = \left(
\begin{array}{cc}
 1 & -\frac{(q-t) u_2}{q \left(u_1-u_2\right)} \\
 0 & 1 \\
\end{array}
\right),
\end{equation}
and
\begin{equation}
  \label{eq:72}
  \mathcal{R}_{\text{ord Mac}} = K^{*} \frac{1}{||\widetilde{M}||^2}\mathcal{R}
  \frac{1}{||\widetilde{M}||^2} K
\end{equation}
The resulting $2\times 2$ block is the same as the block appearing in
the standard trigonometric $\mathcal{R}$-matrix:
\begin{equation}
  \label{eq:69}
  \mathcal{R}_{\text{ord Mac}} = \left(
\begin{array}{cc}
 \frac{q u_1 \left(u_1-u_2\right)}{u_2 \left(q u_1-t u_2\right)} & \frac{(q-t)
   u_1}{q u_1-t u_2} \\
 \frac{(q-t) u_1^2}{u_2 \left(q u_1-t u_2\right)} & \frac{t u_1
   \left(u_1-u_2\right)}{u_2 \left(q u_1-t u_2\right)} \\
\end{array}
\right)
\end{equation}
Again the digression on the triangular decomposition is relevant
here. The only difference is that the geometric interpretation now
lies in equivariant $K$-theory of the instanton moduli
space. Otherwise, the comparison with the ``wall
$\mathcal{R}$-matrices'' as in Eq.~(\ref{eq:62}) is still valid and
the resulting decomposition also gives the relation with the spectral
duality transformation $\mathcal{S}$ as in Eq.~(\ref{eq:22}). Also, the
DIM $\mathcal{R}$-matrix provides the reflection matrix for the
$q$-deformed Virasoro algebra.

\paragraph{More strands.}
\label{sec:more-strands}
Again the discussion here is exactly parallel to the previous section,
only the formulas are somewhat larger. The generalized Macdonald
Hamiltonian for $N$ strands is given by
\begin{multline}
  \label{eq:87}
  {\cal H}_1 = \rho_{u_1} \otimes \cdots \otimes \rho_{u_N} \Delta
  (x^{+}(z))
  = \oint \frac{dz}{z} \sum_{i=1}^N u_i \Lambda_i (z) =\\
  =\oint \frac{dz}{z} \Biggl[ u_1 \exp \left( \sum_{n \geq 1}
    \frac{1-t^{-n}}{n} p^{(1)}_n z^{-n} \right) \exp \left( \sum_{n
      \geq 1} \frac{1-q^n}{n} z^n \frac{\partial}{\partial p^{(1)}_n}
  \right)+\\
  + u_2 \exp \left( \sum_{n \geq 1} \frac{1-t^{-n}}{n} \left( (1 -
      t^n/q^n) p^{(1)}_n + p^{(2)}_n(q/t)^{-n/2} \right) z^{-n}
  \right) \exp \left( \sum_{n \geq 1} \frac{1-q^n}{n} z^n (q/t)^{n/2}
    \frac{\partial}{\partial p^{(2)}_n} \right)+\\
  + u_3 \exp \left( \sum_{n \geq 1} \frac{1-t^{-n}}{n} \left( (1 -
      t^n/q^n)\left( p^{(1)}_n + (q/t)^{-1/2} p^{(2)}_n \right) +
      p^{(3)}_n(q/t)^{-n} \right) z^{-n} \right) \exp \left( \sum_{n
      \geq 1} \frac{1-q^n}{n} z^n (q/t)^n
    \frac{\partial}{\partial p^{(3)}_n} \right) +\\
  \ldots + u_N \exp \left( \sum_{n \geq 1} \frac{1-t^{-n}}{n} \left(
      (1 - t^n/q^n)\left( p^{(1)}_n + (q/t)^{-1/2} p^{(2)}_n + \ldots
        + (q/t)^{(2-N)n/2} p^{(M-1)}_n\right) +
      p^{(N)}_n(q/t)^{(1-N)/2} \right) z^{-n} \right)\times\\
  \times \exp \left( \sum_{n \geq 1} \frac{1-q^n}{n} z^n
    (q/t)^{(1-N)/2} \frac{\partial}{\partial p^{(3)}_n} \right)
  \Biggr]
\end{multline}
and the generalized Macdonald polynomials are defined as its
eigenfunctions:
\begin{gather}
  \label{eq:89}
  \mathcal{H}_1 \widetilde{M}_{A_1\ldots A_N} = \kappa_{A_1 \ldots
    A_N} \widetilde{M}_{A_1\ldots A_N}\\
  \kappa_{A_1 \ldots A_N} = \sum_{a=1}^N u_a \sum_{i\geq 1}
  q^{A_{a,i}} t^{-i}
\end{gather}

An explicit computation of the $\mathcal{R}$-matrix for three strands is
performed in Appendices~\ref{sec:q-t-deformed} (first level)
and~\ref{sec:q-t-deformed-1} (second level). In the basis of ordinary
Macdonald polynomials, the $\mathcal{R}$ matrices $\mathcal{R}_{ij}$
act only on the $i$-th and $j$-th representations in the tensor
product, just as for any standard integrable system. The Yang-Baxter
equation is also satisfied, as shown in
Appendix~\ref{sec:q-t-deformed}.

\bigskip

Thus, in this $\mathcal{R}$ matrix section, we demonstrated that
both the DIM $\mathcal{R}$-matrix and
its affine Yangian limit can be easily computed for the horizontal Fock
representations using the generalized Jack or Macdonald polynomials. The
resulting $\mathcal{R}$-matrices have usual properties and resemble
the standard rational and trigonometric $\mathcal{R}$-matrices. They are
also related to the ($q$-)Virasoro reflection matrices and can be understood
as the intersection of stable envelopes in the cohomology or
$K$-theory of the instanton moduli spaces.

In the next section, we show that the transfer matrices or the Lax
operators permuted by the DIM $\mathcal{R}$ matrix, can be understood
as refined topological string amplitudes on resolved conifold. We assume that all correlators are normalized in such a way that, for the empty diagrams, the averages are identities.

\section{$\mathcal{RTT}$ relations in the toric diagram}
\label{sec:rtt-relations}
In this section, we prove that the $\mathcal{R}$-matrix \emph{permutes}
the basic building blocks of the balanced toric web. These basic
building blocks are resolved conifolds with Young diagrams placed on
each external line:
\begin{equation}
  \label{eq:14}
  \mathcal{T}_{AB}^{RP} (Q,u,z)
  = \parbox{2.5cm}{\includegraphics[width=2.5cm]{conifold-crop}}
  =\Bigl(\langle s_A, Qu | \otimes \langle M_R^{|}|\Bigr) \mathcal{T}(Q|z)\Bigl(| s_B,u
  \rangle \otimes |M_P^{|}\rangle\Bigr)=
  \langle s_A, Qu |  \Psi_P (Q z) \Psi^{*}_R (z) | s_B,u \rangle,
\end{equation}
where $\Psi$ and $\Psi^{*}$ are the intertwiners of DIM algebra \cite{AFS,DIM1}, $|
s_A ,u \rangle$ denote the basis of Schur functions in the horizontal
Fock space, $|M_R^{|}\rangle$ denote the basis of ordinary Macdonald
polynomials in the vertical Fock space (hence, the sign~$^{|}$) and
$\Psi_P$ denotes the matrix element of $\Psi$ for the Macdonald polynomial
$M_P$ on the vertical leg. Such building blocks allow us to construct
an arbitrary balanced networks as shown in~\cite{DIM1}.

\subsection{Trivial diagrams on vertical legs}
\label{sec:rtt-relations-with}
Before considering the most general $\mathcal{RTT}$ relation, let us give the
proof in the simplified case, where some of the external diagrams are
empty. The main ideas of the proof are similar to the general case,
which requires one additional observation.

The $\mathcal{RTT}$ relations for the conifold building blocks look very similar
to the $\mathcal{RTT}$ relations in any integrable system and can be drawn as
follows:
\begin{equation}
  \label{eq:15}
  \parbox{14cm}{\includegraphics[width=14cm]{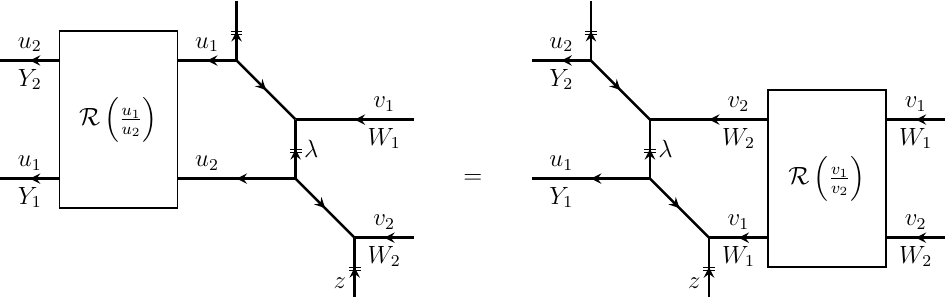}}
\end{equation}
Here $\mathcal{R}$, drawn as a box acts on the tensor product of the
horizontal Fock modules corresponding to the horizontal legs. The
preferred direction is vertical. Two horizontal modules are
intertwined with one vertical by the combination of topological
vertices. Equivalently, in the algebraic form, we have:
\begin{equation}
  \label{eq:16}
  \mathcal{R}\left( \frac{u_1}{u_2} \right) \sum_{\lambda} ||M_{\lambda}||^{-2}
  \begin{smallmatrix}
    \Psi_{\varnothing} \left(\frac{zu_1u_2}{v_1v_2}\right) \Psi^{*}_{\lambda} \left(\frac{zu_2}{v_2}\right)\\
    \otimes\\
    \Psi_{\lambda} \left(\frac{zu_2}{v_2}\right) \Psi^{*}_{\varnothing}(z)
\end{smallmatrix}
=
\sum_{\lambda} ||M_{\lambda}||^{-2}
  \begin{smallmatrix}
    \Psi_{\varnothing} \left(\frac{zu_1u_2}{v_1v_2}\right)
      \Psi^{*}_{\lambda} \left(\frac{zu_1}{v_1}\right)\\
    \otimes\\
    \Psi_{\lambda} \left(\frac{zu_1}{v_1}\right)
        \Psi^{*}_{\varnothing}(z)
\end{smallmatrix}
  \mathcal{R}\left( \frac{v_1}{v_2} \right)
\end{equation}
Notice that the Young diagrams on the vertical external legs are chosen
empty. This is the simplification that we use in this subsection and lift in
the next one.

We will use the following trick, which renders the $\mathcal{RTT}$ relations
almost trivial. We rotate the preferred direction in the diagram from
vertical to horizontal with the help of the automorphism
$\mathcal{S}$. This corresponds to the change of basis in the tensor
product of horizontal representations from that of Schur functions
$|s_{Y_1},u_1\rangle |s_{Y_2},u_2\rangle$ to the generalized Macdonald
polynomials $|M_{Y_1 Y_2}(u_1,u_2|q,t) \rangle$ (without tilde, i.e.\
not specially normalized). In this new basis, the $\mathcal{R}$ matrix
acts simply by permuting the strands (though, as we learned in
the previous section, depending on the normalization of the basis
vectors an additional constant might arise). Thus, in this basis, one
gets the following relation (we moved two $\mathcal{R}$-matrices
to the r.h.s. of Eq.~(\ref{eq:16})):
\begin{equation}
  \label{eq:77}
    \parbox{14cm}{\includegraphics[width=14cm]{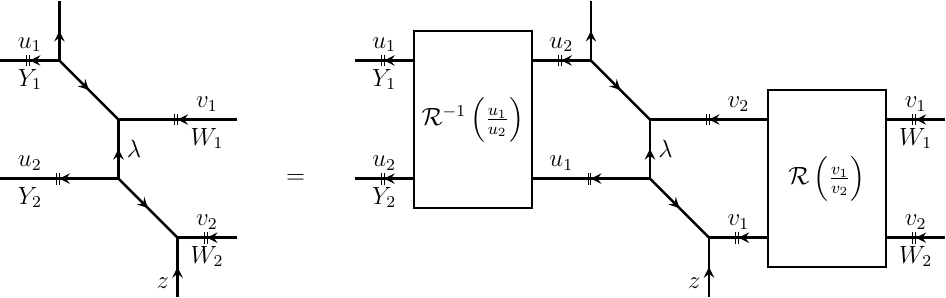}}
\end{equation}
or, algebraically,
\begin{multline}
  \label{eq:79}
  \left\langle M_{Y_1Y_2}\left( \frac{u_1}{u_2} \right)\Bigg| \sum_{\lambda} ||M_{\lambda}||^{-2}
  \begin{smallmatrix}
    \Psi_{\varnothing} \left(\frac{zu_1u_2}{v_1v_2}\right)
    \Psi^{*}_{\lambda} \left(\frac{zu_2}{v_2}\right)\\
    \otimes\\
    \Psi_{\lambda} \left(\frac{zu_2}{v_2}\right)
    \Psi^{*}_{\varnothing}(z)
  \end{smallmatrix}
  \Bigg| M_{W_1 W_2}\left( \frac{v_1}{v_2} \right) \right\rangle
  =\\
  =  \left\langle M_{Y_1Y_2}\left( \frac{u_1}{u_2} \right)\Bigg| \mathcal{R}\left( \frac{u_1}{u_2} \right)^{-1}
  \sum_{\lambda} ||M_{\lambda}||^{-2}
  \begin{smallmatrix}
    \Psi_{\varnothing} \left(\frac{zu_1u_2}{v_1v_2}\right)
    \Psi^{*}_{\lambda} \left(\frac{zu_1}{v_1}\right)\\
    \otimes\\
    \Psi_{\lambda} \left(\frac{zu_1}{v_1}\right)
        \Psi^{*}_{\varnothing}(z)
\end{smallmatrix}
\mathcal{R}\left( \frac{v_1}{v_2} \right)  \Bigg| M_{W_1 W_2}\left(
  \frac{v_1}{v_2} \right) \right\rangle
\end{multline}
Strictly speaking, we should have also changed the basis in the
vertical legs from the basis of Macdonald polynomials to that of Schur
functions. However, the change of the basis in the vertical legs does
not make any difference, since the external diagrams are empty, and
the internal ones are summed over. Let us now use our definition of the
$\mathcal{R}$ matrix~\eqref{tildeMvsM} and the definition of the
opposite generalized Macdonald polynomials~\eqref{eq:81} to transform
the r.h.s.\ Eq.~\eqref{eq:79}:
\begin{multline}
  \label{eq:82}
  \left\langle M_{Y_1Y_2}\left( \frac{u_1}{u_2} \right)\Bigg| \mathcal{R}\left( \frac{u_1}{u_2} \right)^{-1}
  \sum_{\lambda} ||M_{\lambda}||^{-2}
  \begin{smallmatrix}
    \Psi_{\varnothing} \left(\frac{zu_1u_2}{v_1v_2}\right)
    \Psi^{*}_{\lambda} \left(\frac{zu_1}{v_1}\right)\\
    \otimes\\
    \Psi_{\lambda}
    \left(\frac{zu_1}{v_1}\right)
        \Psi^{*}_{\varnothing}(z)
\end{smallmatrix}
\mathcal{R}\left( \frac{v_1}{v_2} \right)  \Bigg| M_{W_1 W_2}\left(
  \frac{v_1}{v_2} \right) \right\rangle =\\
= \frac{N_{Y_2Y_1}\left( \frac{u_2}{u_1} \right)}{N_{Y_1 Y_2}\left( \frac{u_1}{u_2} \right)}  \left\langle M_{Y_2Y_1}\left( \frac{u_2}{u_1} \right)\Bigg|\sum_{\lambda} ||M_{\lambda}||^{-2}
  \begin{smallmatrix}
    \Psi_{\lambda} \left(\frac{zu_1}{v_1}\right)
    \Psi^{*}_{\varnothing}(z)\\
    \otimes\\
    \Psi_{\varnothing} \left(\frac{zu_1u_2}{v_1v_2}\right)
        \Psi^{*}_{\lambda} \left(\frac{zu_1}{v_1}\right)
\end{smallmatrix}\Bigg| M_{W_2 W_1}\left( \frac{v_2}{v_1} \right) \right\rangle \frac{N_{W_2W_1}\left( \frac{v_2}{v_1} \right)}{N_{W_1 W_2}\left( \frac{v_1}{v_2} \right)}
\end{multline}
Notice the change in ordering of the tensor product due to the
exchange of $p_n$ and $\bar{p}_n$ in $M^{\mathrm{op}}$. It remains to
prove the identity between l.h.s.\ of Eq.~\eqref{eq:79} and r.h.s.\ of
Eq.~\eqref{eq:82}. Both expressions are matrix elements of the
$\mathcal{T}$-operators in the basis of generalized Macdonald
polynomials.

We employ a very nice property of the generalized Macdonald basis. In
this basis, the matrix elements of the product of two
$\mathcal{T}$-matrices are explicitly computable and given by the Nekrasov
functions. As shown in~\cite{MZDecompNekr}, the corresponding matrix
model averages factorize and the answer can be schematically written
as follows (we again omit the prefactors, which cancel in the both sides
of the $\mathcal{RTT}$ relations):
\begin{equation}
  \label{eq:78}
 \left\langle M_{Y_1Y_2}\left( \frac{u_1}{u_2} \right)\Bigg| \sum_{\lambda} ||M_{\lambda}||^{-2}
  \begin{smallmatrix}
    \Psi_{\varnothing} \left(\frac{zu_1u_2}{v_1v_2}\right)
        \Psi^{*}_{\lambda} \left(\frac{zu_2}{v_2}\right)\\
    \otimes\\
    \Psi_{\lambda} \left(\frac{zu_2}{v_2}\right)
        \Psi^{*}_{\varnothing}(z)
  \end{smallmatrix} \Bigg| M_{W_1 W_2}\left( \frac{v_1}{v_2} \right)
\right\rangle \sim \frac{z_{\mathrm{bifund}}^{(q,t)}\left([Y_1, Y_2],[W_1,
    W_2], \frac{u_1}{u_2}, \frac{v_1}{v_2}, \frac{u_1u_2}{v_1
      v_2}\right)}{G_{Y_1,Y_2}\left( \frac{u_1}{u_2} \Big| q,t\right)
  G_{W_1,W_2} \left( \frac{v_1}{v_2} \Big| q,t \right)}
\end{equation}
The $G$ factors on the both sides of the $\mathcal{RTT}$ relations in the
denominator cancel with the normalization factors $N_{Y_1Y_2}$
and~(\ref{eq:79}) reduces to the elementary identity for the
bifundamental Nekrasov functions \cite{Nek,NekOk}:
\begin{equation}
  \label{eq:80}
  z_{\mathrm{bifund}}^{(q,t)}([Y_1, Y_2],[W_1, W_2],Q_u, Q_v,M) =   z_{\mathrm{bifund}}^{(q,t)}([Y_2, Y_1],[W_2, W_1],Q_u^{-1}, Q_v^{-1},M)
\end{equation}
Thus, we proved the $\mathcal{RTT}$ relation~\eqref{eq:16} for the empty
diagrams on the vertical legs.

\subsection{Arbitrary diagrams on vertical legs}
\label{sec:arbitr-diagr-vert}
We now generalize our proof of the $\mathcal{RTT}$ relations to the
case of arbitrary states in the vertical representations. To this end,
we use explicit expressions for the DIM intertwiners acting in the
horizontal Fock module with the spectral parameter $z$
(see~\cite{AFS,DIM1} for details):
\begin{gather}
  \label{eq:75}
  \Psi_{\lambda}(v) = (-vz)^{|\lambda|}
  \frac{q^{n(\lambda^{\mathrm{T}})}}{ C_{\lambda}(q,t)} \exp \left(
    \sum_{n \geq 1} \frac{1}{n} \frac{1-t^n}{1-q^n} \sum_{i \geq 1}
    q^{n \lambda_i} t^{-in} v^n a_{-n} \right) \exp \left( - \sum_{n
      \geq 1} \frac{1}{n} \frac{1-t^n}{1-q^n}\sum_{i \geq 1} q^{-n\lambda_i} t^{in}
    \left( \frac{q}{t} \right)^n
    v^{-n} a_n    \right),\\
  \Psi^{*}_{\mu}(u) = \left(-\frac{uz}{q}\right)^{-|\mu|}
  \frac{q^{n(\mu^{\mathrm{T}})}}{f_{\mu} C_{\mu}(q,t)} \exp \left(
    -\sum_{n\geq 1} \frac{1}{n} \frac{1-t^n}{1-q^n} \sum_{i \geq 1}
    q^{n \mu_i} t^{-ni} \left( \frac{t}{q} \right)^{\frac{n}{2}} u^n
    a_{-n} \right)\times \nn\\
  \times \exp \left(
    \sum_{n\geq 1} \frac{1}{n} \frac{1-t^n}{1-q^n} \sum_{i \geq 1}
    q^{-n \mu_i} t^{ni} \left( \frac{q}{t} \right)^{\frac{n}{2}} u^{-n} a_n \right)
\end{gather}
where
\begin{equation}
  f_\lambda = \prod_{(i,j)\in\lambda} (-q^{j-1/2} t^{1/2-i}),
    \qquad n(\lambda^T)=\sum_{(i,j)\in\lambda} (j-1)
\end{equation}
The combination of intertwiners entering Eq.~(\ref{eq:14}) is then
given by:
\begin{multline}
  \label{eq:76}
  \langle M_\lambda^{|}|\mathcal{T}|M_\mu^{|}\rangle= \Psi_{\lambda}(v)
  \Psi^{*}_{\mu}(u) = W_{\lambda \mu} (z,u,v) \exp \left( \sum_{n \geq
      1} \frac{1}{n} \frac{1-t^n}{1-q^n} \sum_{i \geq 1} \left( q^{n
        \lambda_i} t^{-in} v^n - q^{n \mu_i} t^{-ni} \left(
        \frac{t}{q} \right)^{\frac{n}{2}} u^n \right) a_{-n}
  \right)\times\\
  \times \exp \left( \sum_{n \geq 1} \frac{1}{n} \frac{1 - t^n}{1 -
      q^n} \sum_{i \geq 1} \left( - q^{-n\lambda_i} t^{in} \left(
        \frac{q}{t} \right)^nv^{-n} + q^{-n\mu_i} t^{ni} \left( \frac{q}{t}
      \right)^{\frac{n}{2}} u^{-n} \right)a_n\right)
\end{multline}
where $W_{\lambda,\mu}(z,u,v) \sim z^{|\lambda|-|\mu|}\Delta^{(q,t)}(u
q^{\lambda} t^{\rho}, v q^{\mu} t^{\rho})^{-1}$ is the scalar
prefactor including the prefactors of $\Psi$ and $\Psi^{*}$ and the
terms from the normal ordering of $\Psi$ and $\Psi^{*}$. We will
henceforth omit this prefactor in our calculations, since it does not
affect the $\mathcal{RTT}$ relations, which are homogeneous in $\mathcal{T}$.

The identity we would like to prove can be represented pictorially as
\begin{equation}
  \label{eq:83}
  \parbox{14cm}{\includegraphics[width=14cm]{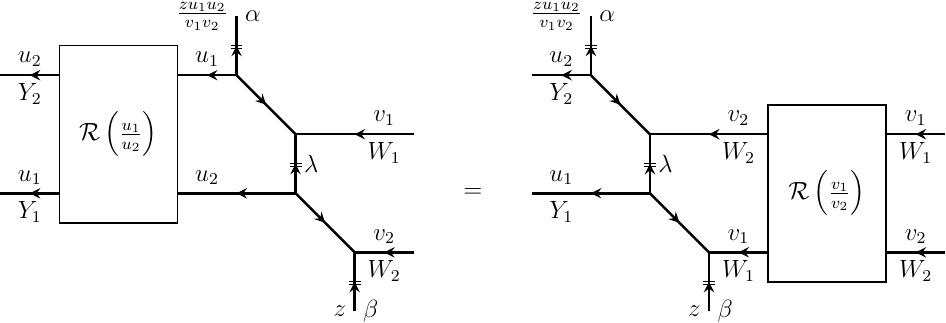}}
\end{equation}
or, algebraically,
\begin{equation}
  \label{eq:84}
  \mathcal{R}\left( \frac{u_1}{u_2} \right) \sum_{\lambda} ||M_{\lambda}||^{-2}
  \begin{smallmatrix}
    \Psi_{\alpha} \left(\frac{zu_1u_2}{v_1v_2}\right)
    \Psi^{*}_{\lambda} \left(\frac{zu_2}{v_2}\right)\\
    \otimes\\
    \Psi_{\lambda} \left(\frac{zu_2}{v_2}\right) \Psi^{*}_{\beta}(z)
\end{smallmatrix}
=
\sum_{\lambda} ||M_{\lambda}||^{-2}
  \begin{smallmatrix}
    \Psi_{\alpha} \left(\frac{zu_1u_2}{v_1v_2}\right)
        \Psi^{*}_{\lambda} \left(\frac{zu_1}{v_1}\right)\\
    \otimes\\
    \Psi_{\lambda} \left(\frac{zu_1}{v_1}\right)
        \Psi^{*}_{\beta}(z)
\end{smallmatrix}
  \mathcal{R}\left( \frac{v_1}{v_2} \right)
\end{equation}
for arbitrary $\alpha$ and $\beta$. We use the same trick as in the
previous subsection and rotate the preferred direction of the diagram
from vertical to horizontal. This makes the $\mathcal{R}$-matrix
diagonal as in Eq.~\eqref{eq:82}. However, now we compute the
resulting amplitude in a different way: we also \emph{rotate} the
whole picture by $\frac{\pi}{2}$ and write down the operator
expression for the matrix elements in the \emph{rotated}
frame. Explicitly, we have
\begin{multline}
  \label{eq:85}
  \left\langle s_{\beta}, z \Bigg|
    \Psi^{*}_{W_2}(v_2)\Psi_{Y_2}(u_{2}) \Psi^{*}_{W_1}(v_1)  \Psi_{Y_1}(u_1) \Bigg|
    s_{\alpha}, \frac{z u_1
      u_2}{v_1 v_2} \right\rangle =\\
  = \frac{N_{Y_2Y_1}\left( \frac{u_2}{u_1} \right)}{N_{Y_1Y_2}\left(
      \frac{u_1}{u_2} \right)} \frac{N_{W_2W_1}\left( \frac{v_2}{v_1}
    \right)}{N_{W_1 W_2}\left( \frac{v_1}{v_2} \right)} \left\langle
    s_{\beta}, z \Bigg|  \Psi^{*}_{W_1}(v_1) \Psi_{Y_1}(u_1)
    \Psi^{*}_{W_2}(v_2) \Psi_{Y_2}(u_2) \Bigg| s_{\alpha}, \frac{z u_1 u_2}{v_1 v_2}
  \right\rangle
\end{multline}
which should be valid for any $\alpha$ and $\beta$, so that these
external states can be dropped. We have thus reduced the $\mathcal{RTT}$
relation to the commutation relation for the $\mathcal{T}$-operators
composed of the DIM intertwiners $\Psi$ and $\Psi^{*}$. We normalize the
product of $\mathcal{T}$-operators using the explicit
expressions~\eqref{eq:76}. We obtain
\begin{equation}
  \label{eq:86}
  \Psi^{*}_{W_2}(v_2) \Psi_{Y_2}(u_{2}) \Psi^{*}_{W_1}(v_1) \Psi_{Y_1}(u_1)
  \sim
  \frac{z_{\mathrm{bifund}}^{(q,t)}\left([Y_1, Y_2],[W_1, W_2],
      \frac{u_1}{u_2}, \frac{v_1}{v_2}, \frac{u_1u_2}{v_1
        v_2}\right)}{G_{Y_1,Y_2}\left( \frac{u_1}{u_2} \Big|
      q,t\right) G_{W_1,W_2} \left( \frac{v_1}{v_2} \Big| q,t \right)}
  : \Psi^{*}_{W_2}(v_2) \Psi_{Y_2}(u_{2}) \Psi^{*}_{W_1}(v_1)  \Psi_{Y_1}(u_1):
\end{equation}
where we have dropped inessential prefactors. This result certainly reduces to Eq.~\eqref{eq:78} for $\alpha = \beta =
\varnothing$, since the normally ordered operators act trivially on the
vacuum. One can now obtain the commutation relation for the
$\mathcal{T}$-operators by normal ordering of the both sides of~\eqref{eq:85}
and using the identity~\eqref{eq:80} for the Nekrasov bifundamental
factor.

Let us recapitulate our main point in this section. We proved the
$\mathcal{RTT}$ relations for the DIM $\mathcal{R}$-matrix and
$\mathcal{T}$-operators constructed from refined topological string
amplitudes on resolved conifold. In the next section, we use these
relations to obtain commuting integrals of motion for our system.

\section{Integrals of motion and compactification}
\label{sec:comm-transf-matr}

Just as in any integrable system, the $\mathcal{RTT}$ relations~(\ref{eq:15}) allow
one to construct a commutative family of operators, integrals of
motion on the Hilbert space of the theory. Those are usually taken
to be traces of $\mathcal{T}$-operators in various representations. In
our case, there are several different ways to write down the integrals
of motion. The first possibility is to take the vacuum matrix element
of a product of $\mathcal{T}$-operators. This gives the closed
string amplitude on the toric Calabi-Yau threefold consisting of the resolved
conifolds. The other way is to compactify the toric diagram, which
gives traces of products $\mathcal{T}$-operators. The resulting amplitude is
given by the matrix model average with the \emph{affine} or
\emph{elliptic} measures depending on the direction of
compactification.

A geometric meaning of the commutativity is in the both cases a
generalization of the flop transition on the Calabi-Yau threefold. The most
basic example of this transition is resolved conifold. The resolution can
be taken in two different ways: either in one direction, or
in the other one. The topological string amplitudes on two resolutions are
related to each other by an analytic continuation in the K\"ahler
parameter $Q$ of the resolution. To switch from one threefold to the other, one has
to replace $Q$ by $Q^{-1}$. In the $\mathcal{RTT}$ relation, a similar
exchange happens and, since the spectral parameters of two legs are
exchanged, their \emph{ratio}, i.e. the corresponding K\"ahler parameter
is reversed. However, the situation is here slightly different, since the toric diagram looks the same after the application of the ${\cal R}$-matrix.

Let us consider the two ways to construct the integrals of motion.

\begin{figure}[h]
  \centering
    \includegraphics[width=12cm]{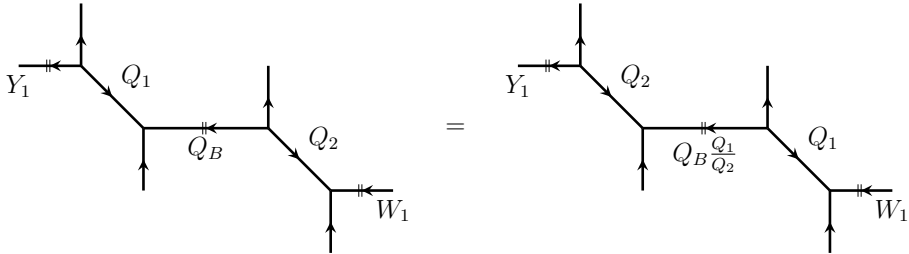}
    \caption{Commutativity of the integrals of motion implies
      relations between amplitudes with different K\"ahler parameters
      on the toric strip geometry.}
  \label{fig:4}
\end{figure}

\paragraph{Vacuum matrix elements.}
\label{sec:vacu-matr-elem}
The simplest example of this form is given by the toric strip geometry
shown in Fig.~\ref{fig:4}. This geometry corresponds to a single chain of
intertwiners with the empty diagrams on all vertical legs. From the
explicit expressions for the intertwiners \cite{DIM1}, one can deduce that the
$\mathcal{T}$ operators indeed commute.

The next step is to glue several strips together. This gives what
was called in~\cite{DIM1} \emph{balanced network}. The
amplitude corresponding to a balanced network can be interpreted as
the partition function of a $5d$ linear quiver gauge theory with zero $\beta$-function. Then,
depending on the duality frame (or preferred direction), the commutativity
of integrals is either related to the action of the Weyl group of
the gauge group or to the \emph{spectral dual} Weyl group. This dual Weyl group
corresponds to the Dynkin diagram of the quiver and permutes the
\emph{gauge coupling constants}. This dual Weyl group
has an interesting interpretation in terms of the AGT dual conformal
block: it exchanges the points and therefore represents some kind
of \emph{a braiding matrix}.

\paragraph{Compactification.}
\label{sec:compactification}
The compactified toric diagram corresponds to elliptically fibered Calabi-Yau
threefolds. Within the geometric engineering approach, such manifolds are
related to gauge theories with adjoint matter, or necklace
quivers. Again, the commutativity of integrals of motion is equivalent to the invariance
under the Weyl group of the corresponding necklace quiver, and thus
permutes the coupling constants of the gauge theory. The spectral dual
interpretation of the resulting amplitude is the partition function of
a $6d$ \emph{linear} quiver gauge theory compactified on a
two-dimensional torus. The AGT relations in this case \cite{AGT5d} give the conformal
block of the $q$-deformed $W$-algebras on torus, or the spherical conformal
block of the \emph{affine} $W$-algebra \cite{MMZ}.

\section{Conclusion}
\label{sec:conclusion}

\begin{figure}[h]
  \centering
    \includegraphics[width=5cm]{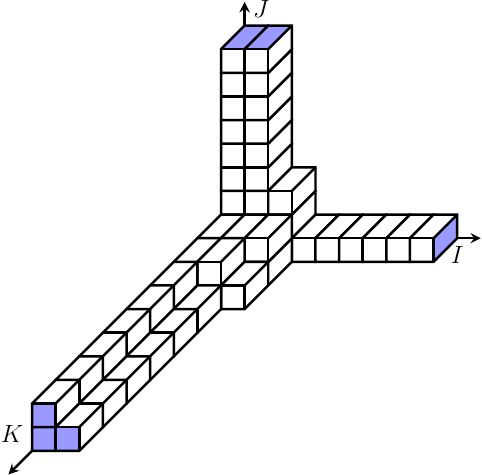}
  \caption{A plane partition (3d Young diagram) with asymptotics given
    by ordinary Young diagrams $I$, $J$ and $K$.}
  \label{fig:3}
\end{figure}

\begin{itemize}
\item \textbf{DIM algebras of higher rank} It would be interesting to
  consider similar construction of the $\mathcal{R}$-matrix for the
  quantum toroidal algebras of higher rank, i.e.\
  $U_{q,t}(\widehat{\widehat{\mathfrak{gl}}}_r)$. However, the
  generalized Macdonald polynomials in this case remain to be
  computed. One of possible difficulties on this way is that
  bosonization involves less trivial free fields {\it a la}
  \cite{GMMOS}.

\item \textbf{Triple-deformed ${\cal R}$-matrix}  If one
  compactifies the toric diagram in the vertical direction, there would emerge
  an ``affinized'' version of the DIM $\mathcal{R}$-matrix. It is plausible
  that this is the $\mathcal{R}$-matrix for the Pagoda algebra \cite{DIM1}, with the additional
  parameter being the compactification radius.

To evaluate this $\mathcal{R}$-matrix, one needs to understand the corresponding
``affine'' generalized Macdonald polynomials. Let us notice that they
exist already for a single horizontal leg, i.e.\ the simplest
example is given by the polynomial labelled by a \emph{single} Young
diagram and depending on \emph{three} parameters
$M_A^{(q,t,\tilde{t})}(p_n)$. For $\tilde{t} \to 0$ one should recover
the ordinary Macdonald polynomials.

\item \textbf{Application to knots} As we already noticed, the DIM algebra can be obtained as a limit of the spherical DAHA
  algebra for infinite number of strands. It is also known that the toric
  knot superpolynomials can be obtained from the action of spherical
  DAHA, or the corresponding DIM (the $SL(2,\mathbb{Z})$ automorphisms
  play a crucial role in these computations). It is natural to assume
  that the $\mathcal{R}$-matrix of DIM should be related to
  computation of superpolynomials. Presumably this
  $\mathcal{R}$-matrix might give the Reshetikhin-Turaev formalism
  behind the Khovanov-Rozansky cohomologies.

\item \textbf{Building representations.} Another important direction
  is to consider more sophisticated representations of the DIM
  algebra. Those can be obtained out of Fock modules by taking tensor
  products, e.g.\ $\mathcal{F}_{u_1}\otimes \cdots \otimes
  \mathcal{F}_{u_k}$. For $u_i$ in general position, these
  representations turn out to be irreducible. The $\mathcal{R}$-matrix
  for these representations is given by the fusion construction, which
  is similar to the known technique for affine quantum algebras. The
  $\mathcal{R}$-matrix for these discrete choices of parameters also
  satisfies the Hecke algebra relations.

However, as in the case of affine quantum algebras, (see, e.g., \cite{Jimbo}) for certain
discrete choice of parameters $u_a$ in \emph{resonance}, i.e.\ for
\begin{equation}
  \label{eq:13}
  u_1 = q^{i_1} t^{-j_1} u_2, \qquad u_2 = q^{i_2} t^{-j_2} u_3, \dots
\end{equation}
with $i_a, j_a \in \mathbb{Z}_{\geq 0}$, one gets invariant subspaces
inside the tensor product arising from degenerate vectors. After
factoring out these subspaces, one gets an irreducible representation
space spanned by a subset of $k$-tuples of Young diagrams obeying
additional \emph{Burge conditions}~\cite{Bershtein-Foda, Foda,Feigin}. These
conditions can be interpreted as the requirement that the $k$-tuple of
Young diagrams combine into a \emph{plane partition} (3d Young
diagram, melting crystal) of width $k$. In general, one can consider 3d
Young diagrams with infinite ``legs'', i.e.\ nontrivial asymptotics along the
coordinate axes, see Fig.~\ref{fig:3}. In the case of the tensor product, the
``vertical'' leg is nontrivial and is determined by the collection of
numbers $\{ i_a, j_a \}$ in the resonance condition~(\ref{eq:13}). To
get a second nontrivial leg, one should consider an infinite
tensor $\lim_{k\to \infty} \bigotimes_{i=1}^k\mathcal{F}_{u_i}$. In
this case, the second asymptotic of the 3d Young diagram is determined
by the asymptotic shape of the last Young diagram $\lim_{k\to \infty}
Y_k$. The resulting representation is called the \emph{MacMahon
  module} and has many nice properties, e.g.\ the $\mathfrak{S}_3$
symmetry exchanging the coordinate axes. Since the MacMahon module is a
subrepresentation of the tensor product, it is, in principle, possible
to obtain the $\mathcal{R}$-matrix for it by the fusion method. The
$\mathcal{R}$-matrix in this case should be related to the one studied
in~\cite{FJMM-Bethe,FJMMRmat}. It would be interesting to see if
this calculation can be made explicitly.

\end{itemize}

\appendix

\section{Explicit expressions for DIM $\mathcal{R}$-matrix}
\label{sec:some-expl-expr}
In this Appendix, we provide explicit expressions for the DIM $\mathcal{R}$-matrix at
the first two levels of Fock representations for two and three
strands. We then demand that the three $\mathcal{R}$-matrices acting on
the three strands, $\mathcal{R}_{12}$, $\mathcal{R}_{23}$ and
$\mathcal{R}_{13}$ act on its own pair of spaces each. This means
that $\mathcal{R}_{12}$ acts exclusively on the first two polynomials in
the basis $M_A(p^{(1)})M_B(p^{(2)})M_C(p^{(3)})$.  This allows us to
find the special normalization constants, which we denote here by
$k^{ij}_l$. They are related to the normalization constants
$N_{AB}(u|q,t)$ featuring in the main text, e.g.\:
\begin{equation}
  \label{eq:64}
  k_2^{12} = \frac{N_{A_1 A_2}\left(\frac{u_1}{u_2}\Big|q,t\right)}{N_{A_2 A_1}\left(\frac{u_2}{u_1}\Big|q,t\right)}
\end{equation}
We also compute the affine Yangian limit of our $\mathcal{R}$-matrix
and verify that it is, indeed, of the form (\ref{eq:58}).

Throughout this Appendix, we use the following notation:
\begin{enumerate}
\item $\Delta$ --- coproduct of DIM algebra
\item $\Delta^{\mathrm{op}}$ --- opposite coproduct
\item $\rho_u$ --- horizontal level one Fock representation of DIM
\item $\rho^{(N)}_{u_1,\ldots,u_N} =
  (\rho_{u_1}\otimes \cdots \otimes \rho_{u_N}) \circ (\Delta \otimes \mathrm{id} \otimes \cdots \otimes \mathrm{id}) \circ \cdots \circ (\Delta \otimes \mathrm{id}) \circ \Delta$
\item $M_{A^{(1)},\ldots,
    A^{(N)}}\left(u_1,\ldots,u_N\Big|q,t\Big|p^{(1)},\ldots,
    p^{(N)}\right)$ --- generalized Macdonald polynomial, i.e.\
  eigenfunctions of $\rho^{(N)}_{u_1,\ldots,u_N}(x^+_0)$. Note that
  they satisfy the following filtration property $M_{AB
    \emptyset}\left(u_1,u_2,u_3\Big|q,t\Big|p^{(1)},p^{(2)},
    p^{(3)}\right) =
  M_{AB}\left(u_1,u_2\Big|q,t\Big|p^{(1)},p^{(2)}\right)$.
\end{enumerate}

\subsection{R-Matrix at level 1}
\label{sec:r-matrix-at}

\subsubsection{$(q,t)$-deformed version}
\label{sec:q-t-deformed}
In this Appendix, we formally write the ${\cal R}$-matrix as $\mathcal{ R} = \sum_i
a_i \otimes b_i$ and set $\mathcal{R}_{12}= \sum_i a_i \otimes b_i
\otimes 1$, $\mathcal{R}_{23}= \sum_i 1 \otimes a_i \otimes b_i$,
$\mathcal{R}_{13}= \sum_i a_i \otimes 1 \otimes b_i$.  In order to
obtain the representation matrix of $\mathcal{R}_{ij}$, we need the
generalized Macdonald polynomials $M_{ABC}$ in $N=3$ case.  The
following are examples of $M_{ABC}$ at level 1:
\begin{equation}
\left(
\begin{array}{c}
 M_{\emptyset,\emptyset,[1]} \\
M_{\emptyset,[1],\emptyset}\\
M_{[1],\emptyset,\emptyset}
\end{array}
\right)
=A(u_1,u_2,u_3)
\left(
\begin{array}{c}
 p_{\emptyset,\emptyset,[1]} \\
p_{\emptyset,[1],\emptyset}\\
p_{[1],\emptyset,\emptyset}
\end{array}
\right)
\end{equation}
\begin{equation}
A(u_1,u_2,u_3):=\left(
\begin{array}{ccc}
 1 & -\frac{(q-t) u_3}{\sqrt{\frac{q}{t}} t (u_2-u_3)} & -\frac{(q-t) u_3 (q u_3-t u_2)}{q t (u_1-u_3) (u_3-u_2)} \\
 0 & 1 & -\frac{(q-t) \sqrt{\frac{q}{t}} u_2}{q (u_1-u_2)} \\
 0 & 0 & 1 \\
\end{array}
\right)
\end{equation}
By definition of the ${\cal R}$-matrix, one has
\begin{equation}
(\Delta^{\mathrm{op}} \otimes \mathrm{id}) \circ \Delta (x^+_0)
= \mathcal{R}_{12}
 ({\Delta} \otimes \mathrm{id}) \circ \Delta (x^+_0)
 \mathcal{R}^{-1}_{12}.
\end{equation}
Thus, $\rho_{u_1u_2u_3}(\mathcal{R}_{12}) M_{ABC}$
is proportional to an eigenfunction of $\rho_{u_1u_2u_3}((\Delta^{\mathrm{op}} \otimes \mathrm{id}) \circ \Delta (x^+_0) )$, where $\rho_{u_1u_2u_3}=\rho_{u_1}\otimes \rho_{u_2} \otimes \rho_{u_3}$.
Its eigenfunctions $M^{(12)}_{DEF}$ are obtained by replacing $\po$ with $\pt$ and
$u_1$ with $u_2$, i.e.,
\begin{align}
& \rho_{u_1u_2u_3}\left( (\Delta^{\mathrm{op}} \otimes \mathrm{id}) \circ \Delta (x^+_0) \right)  M^{(12)}_{ABC} = e_{ABC} M^{(12)}_{ABC}, \\
& M^{(12)}_{ABC}:=M_{BAC}(u_2,u_1,u_3|q,t|\pt,\po,p^{(3)}).
\end{align}
Then, the eigenvalues $e_{ABC}$ are the same as those for $M_{ABC}$.
Therefore,
if we set the matrix
\begin{equation}
B^{(12)}:=
\left(\begin{array}{ccc}
 k^{(12)}_1 & 0 & 0 \\
 0 &  k^{(12)}_2 & 0 \\
 0 & 0 &  k^{(12)}_3 \\
\end{array}\right)
\left(\begin{array}{ccc}
 1 & 0 & 0 \\
 0 & 0 & 1 \\
 0 & 1 & 0 \\
\end{array}\right)
A(u_2,u_1,u_3)
\left(\begin{array}{ccc}
 1 & 0 & 0 \\
 0 & 0 & 1 \\
 0 & 1 & 0 \\
\end{array}\right)
A^{-1}(u_1,u_2,u_3),
\end{equation}
then the representation matrix of $\rho_{u_1u_2u_3}(\mathcal{R}_{12}) $
in the basis of generalized Macdonald polynomials
is the transposed matrix of $B^{(12)}$:
\begin{equation}
\rho_{u_1u_2u_3}(\mathcal{R}_{12})
\left(\begin{array}{ccc}
 M_{\emptyset,\emptyset,[1]} & M_{\emptyset,[1],\emptyset} & M_{[1]\emptyset,\emptyset} \\
\end{array}\right)
=
\left(\begin{array}{ccc}
 M_{\emptyset,\emptyset,[1]} & M_{\emptyset,[1],\emptyset} & M_{[1]\emptyset,\emptyset} \\
\end{array}\right)
{}^t B^{(12)},
\end{equation}
where $k^{(12)}_i$ are the proportionality constants between $M^{(12)}_{ABC}$ and $\rho (\mathcal{R}_{12})(M_{ABC})$.

In the same way, from the formula
\begin{equation}
(\mathrm{id} \otimes \Delta^{\mathrm{op}}) \circ \Delta (x^+_0)
= \mathcal{R}_{23}
 ({\Delta} \otimes \mathrm{id}) \circ \Delta (x^+_0)
 \mathcal{R}^{-1}_{23},
\end{equation}
the representation matrix of $\rho_{u_1u_2u_3}(\mathcal{R}_{23}) $ is
\begin{equation}
\rho_{u_1u_2u_3}(\mathcal{R}_{23})
\left(\begin{array}{ccc}
 M_{\emptyset,\emptyset,[1]} & M_{\emptyset,[1],\emptyset} & M_{[1]\emptyset,\emptyset} \\
\end{array}\right)
=
\left(\begin{array}{ccc}
 M_{\emptyset,\emptyset,[1]} & M_{\emptyset,[1],\emptyset} & M_{[1]\emptyset,\emptyset} \\
\end{array}\right)
{}^t B^{(23)},
\end{equation}
where
\begin{equation}
B^{(23)}:=
\left(\begin{array}{ccc}
 k^{(23)}_1 & 0 & 0 \\
 0 &  k^{(23)}_2 & 0 \\
 0 & 0 &  k^{(23)}_3 \\
\end{array}\right)
\left(\begin{array}{ccc}
 0 & 1 & 0 \\
 1 & 0 & 0 \\
 0 & 0 & 1 \\
\end{array}\right)
A(u_1,u_3,u_2)
\left(\begin{array}{ccc}
 0 & 1 & 0 \\
 1 & 0 & 0 \\
 0 & 0 & 1 \\
\end{array}\right)
A^{-1}(u_1,u_2,u_3).
\end{equation}

The constants $k^{(ij)}_l$ are determined as follows.
At first,
since the scalar multiples of ${\cal R}$-matrices are also ${\cal R}$-matrices,
we can normalize $k^{(12)}_1=k^{(23)}_3=1$.
This means that $\mathcal{R} (1\otimes1)=1\otimes 1$.
Now
we consider the basis change from $M_{ABC}$ to power sum symmetric functions:
\begin{equation}
\widetilde{B}^{(ij)}:= {}^tA(u_1,u_2,u_3)\, {}^tB^{(ij)}\, {}^tA^{-1}(u_1,u_2,u_3).
\end{equation}
Then $\widetilde{B}^{(ij)}$ have the following form
\begin{equation}
\widetilde{B}^{(12)}=
\left(\begin{array}{ccc}
 1 & 0 & 0 \\
 b^{(12)}_2 & * & * \\
 b^{(12)}_3 & * & * \\
\end{array}\right),
\quad
\widetilde{B}^{(23)}=
\left(\begin{array}{ccc}
 * & * & 0 \\
 * & * & 0 \\
 b^{(23)}_1 & b^{(23)}_2 & 1 \\
\end{array}\right),
\end{equation}
where $b^{(ij)}_n$ are functions of $k^{(ij)}_l$.
Since when $\mathcal{R}_{12}$ acts to $p^{(3)}_1$, the
variables $p^{(1)}_1$ and $p^{(2)}_1$ must not appear,
one gets the equations $b^{(12)}_2=b^{(12)}_3=0$.
Similarly, $b^{(23)}_1=b^{(23)}_2=0$.
By solving these equations,
one can see that
\begin{align}
&
k^{(12)}_1=1,\quad
k^{(12)}_2=-\frac{\sqrt{\frac{q}{t}} (q u_2-t u_1)}{q (u_1-u_2)},\quad
k^{(12)}_3=\frac{t (u_1-u_2) \sqrt{\frac{q}{t}}}{q u_1-tu_2}, \\
&
k^{(23)}_1=\frac{\sqrt{\frac{q}{t}} (t u_2-q u_3)}{q (u_2-u_3)},\quad
k^{(23)}_2=\frac{t (u_2-u_3) \sqrt{\frac{q}{t}}}{q u_2-tu_3},\quad
k^{(23)}_3=1.
\end{align}
In this way, one obtains an explicit expression of the ${\cal R}$-Matrix at level 1
\begin{equation}
\widetilde{B}^{(12)}=
\left(
\begin{array}{ccc}
 1 & 0 & 0 \\
 0 & \frac{\sqrt{\frac{q}{t}} t (u_1-u_2)}{q u_1-t u_2} & \frac{(q-t) u_1}{q u_1-t u_2} \\
 0 & \frac{(q-t) u_2}{q u_1-t u_2} & \frac{\sqrt{\frac{q}{t}} t (u_1-u_2)}{q u_1-t u_2} \\
\end{array}
\right),
\quad
\widetilde{B}^{(23)}=
\left(
\begin{array}{ccc}
 \frac{\sqrt{\frac{q}{t}} t (u_2-u_3)}{q u_2-t u_3} & \frac{(q-t) u_2}{q u_2-t u_3} & 0 \\
 \frac{(q-t) u_3}{q u_2-t u_3} & \frac{\sqrt{\frac{q}{t}} t (u_2-u_3)}{q u_2-t u_3} & 0 \\
 0 & 0 & 1 \\
\end{array}
\right)
\end{equation}
Thus,
using the symmetry w.r.t. $p^{(i)}$ at different $i$,
one also gets the representation matrix of $\rho_{u_1u_2u_3}(\mathcal{R}_{13})$
\begin{equation}
\widetilde{B}^{(13)}=
\left(
\begin{array}{ccc}
 \frac{\sqrt{\frac{q}{t}} t (u_1-u_3)}{q u_1-t u_3} & 0 & \frac{(q-t) u_1}{q u_1-t u_3} \\
 0 & 1 & 0 \\
 \frac{(q-t) u_3}{q u_1-t u_3} & 0 & \frac{\sqrt{\frac{q}{t}} t (u_1-u_3)}{q u_1-t u_3} \\
\end{array}
\right).
\end{equation}
Indeed, one can check that they satisfy the Yang-Baxter equation
\begin{equation}
\widetilde{B}^{(12)}\widetilde{B}^{(13)}\widetilde{B}^{(23)}=\widetilde{B}^{(23)}\widetilde{B}^{(13)}\widetilde{B}^{(12)}.
\end{equation}
Of course, the same equations for $B^{(ij)}$ also follow.
Incidentally,
\begin{equation}
{}^tB^{(12)}=
\left(
\begin{array}{ccc}
 1 & 0 & 0 \\
 \frac{u_3 (q-t) \left(-t u_1 \sqrt{\frac{q}{t}}+q u_3 \sqrt{\frac{q}{t}}+q u_1-q u_3\right)}{q t (u_1-u_3)
   (u_2-u_3) \sqrt{\frac{q}{t}}} & -\frac{\sqrt{\frac{q}{t}} (q u_2-t u_1)}{q (u_1-u_2)} & \frac{u_1
   (q-t)}{q u_1-t u_2} \\
x & -\frac{u_2 (q-t) (q
   u_2-t u_1)}{q t (u_1-u_2)^2} & \frac{\sqrt{\frac{q}{t}} \left(q^2 u_1 u_2+q t u_1^2+q t u_2^2-4 q t
   u_1 u_2+t^2 u_1 u_2\right)}{q (u_1-u_2) (q u_1-t u_2)} \\
\end{array}
\right)
\end{equation}
{\small
\begin{equation}
x= \frac{u_3 (q-t) \left(q^2 u_2 u_3-t^2 u_2^2 \sqrt{\frac{q}{t}}+t^2 u_2 u_3 \sqrt{\frac{q}{t}}+q t u_2^2+q
   t u_1 u_2 \sqrt{\frac{q}{t}}-2 q t u_1 u_2-q t u_1 u_3 \sqrt{\frac{q}{t}}+q t u_1 u_3-2 q t u_2
   u_3+t^2 u_1 u_2\right)}{q t^2 (u_1-u_2) (u_1-u_3) (u_2-u_3) \sqrt{\frac{q}{t}}}
\end{equation}
}
\begin{equation}
{}^tB^{(23)}=
\left(
\begin{array}{ccc}
 \frac{\sqrt{\frac{q}{t}} (t u_2-q u_3)}{q (u_2-u_3)} & \frac{u_2 (q-t)}{q u_2-t u_3} & 0 \\
 \frac{u_3 (q-t) (t u_2-q u_3)}{q t (u_2-u_3)^2} & \frac{\sqrt{\frac{q}{t}} \left(q^2 u_2 u_3+q t u_2^2+q
   t u_3^2-4 q t u_2 u_3+t^2 u_2 u_3\right)}{q (u_2-u_3) (q u_2-t u_3)} & 0 \\
 -\frac{u_3 (q-t) \sqrt{\frac{q}{t}} \left(t u_1 \sqrt{\frac{q}{t}}-t u_2 \sqrt{\frac{q}{t}}-q u_1+t
   u_2\right) (t u_2-q u_3)}{q^2 t (u_1-u_2) (u_1-u_3) (u_2-u_3)} &
 y & 1 \\
\end{array}
\right)
\end{equation}
{\small
\begin{equation}
y=\frac{u_2 (q-t) \left(q^2 u_1
   u_3-t^2 u_3^2 \sqrt{\frac{q}{t}}+t^2 u_1 u_3 \sqrt{\frac{q}{t}}+q t u_3^2-q t u_1 u_2
   \sqrt{\frac{q}{t}}+q t u_1 u_2-2 q t u_1 u_3+q t u_2 u_3 \sqrt{\frac{q}{t}}-2 q t u_2 u_3+t^2 u_2
   u_3\right)}{q t (u_1-u_2) (u_1-u_3) (q u_2-t u_3)}
\end{equation}
}
We do not write down the matrix $B^{(13)}$, since it is too complicated.
The representation matrix of $(\rho_{u_1} \otimes \rho_{u_2}) (\mathcal{R})$ is the $2 \times 2$ matrix block at the lower right corner of $\ {}^tB^{(12)}$
\begin{equation}
(\rho_{u_1} \otimes \rho_{u_2}) (\mathcal{R})
=
\left(
\begin{array}{cc}
 -\frac{\sqrt{\frac{q}{t}} (q u_2-t u_1)}{q (u_1-u_2)} & \frac{(q-t) u_1}{q u_1-t u_2} \\
 -\frac{(q-t) u_2 (q u_2-t u_1)}{q t (u_1-u_2)^2} & \frac{\sqrt{\frac{q}{t}} \left(u_1 u_2 q^2+t u_1^2 q+t
   u_2^2 q-4 t u_1 u_2 q+t^2 u_1 u_2\right)}{q (u_1-u_2) (q u_1-t u_2)} \\
\end{array}
\right)
\end{equation}

\subsubsection{$\beta$-deformed version}
\label{sec:beta-deform-vers}
The generalized Macdonald polynomials are reduced to the generalized Jack polynomials in the limit $q \rightarrow 1$ ($t=q^{\beta}, u_i=q^{u'_i}$) (hereafter in this paragraph we substitute $u'_i$ by $u_i$).
Hence,
the $\beta$-deformed version of ${\cal R}$-matrix $\mathcal{R}^{(\beta)}$ is immediately obtained from the results of the last paragraph.
For example, for the representation $\rho_{u_1}\otimes \rho_{u_2}$ and in the basis of generalized Jack polynomials,
\begin{equation}
\mathcal{R}^{(\beta)}=
\left(
\begin{array}{cc}
 k_2 & k_3 \eta \\
 k_2 \eta & k_3 (1+\eta^2) \\
\end{array}
\right).
\end{equation}
Here
\begin{equation}
k_2 = \lim_{q\rightarrow 1} k_2^{(12)}=\frac{u_1-u_2-1+\beta}{u_1-u_2}, \quad
k_3 = \lim_{q\rightarrow 1} k_3^{(12)}=\frac{u_1-u_2}{u_1-u_2+1-\beta}, \quad
\eta = \frac{1-\beta}{u_2-u_1},
\end{equation}
and the generalized Jack polynomials are
\begin{equation}
J_{\emptyset,[1]} = p^{(2)}_1 - \eta  p^{(1)}_1,
J_{[1],\emptyset} = p^{(1)}_1.
\end{equation}
Then,
\begin{equation}
\mathcal{R}^{(\beta)}_{12}=
\left(
\begin{array}{ccc}
 1 & 0 & 0 \\
 -\frac{(\beta -1)^2}{(u_1-u_3) (u_3-u_2)} & \frac{\beta +u_1-u_2-1}{u_1-u_2} & \frac{\beta -1}{\beta
   -u_1+u_2-1} \\
 -\frac{(\beta -1)^2 (\beta +u_1-u_2-1)}{(u_1-u_2) (u_1-u_3) (u_2-u_3)} & -\frac{(\beta -1) (\beta
   +u_1-u_2-1)}{(u_1-u_2)^2} & \frac{\beta ^2-2 \beta +u_1^2+u_2^2-2 u_1 u_2+1}{(u_1-u_2) (-\beta
   +u_1-u_2+1)} \\
\end{array}
\right),
\end{equation}
\begin{equation}
\mathcal{R}^{(\beta)}_{23}=
\left(
\begin{array}{ccc}
 \frac{\beta +u_2-u_3-1}{u_2-u_3} & \frac{\beta -1}{\beta -u_2+u_3-1} & 0 \\
 -\frac{(\beta -1) (\beta +u_2-u_3-1)}{(u_2-u_3)^2} & \frac{\beta ^2-2 \beta +u_2^2+u_3^2-2 u_2
   u_3+1}{(u_2-u_3) (-\beta +u_2-u_3+1)} & 0 \\
 \frac{(\beta -1)^2 (\beta +u_2-u_3-1)}{(u_1-u_2) (u_1-u_3) (u_2-u_3)} & \frac{(\beta -1)^2 (\beta
   +u_2-u_3-1)}{(u_1-u_2) (u_1-u_3) (\beta -u_2+u_3-1)} & 1 \\
\end{array}
\right).
\end{equation}
In the basis of power sum symmetric functions,
\begin{equation}
\mathcal{R}^{(\beta)}=
\left(
\begin{array}{cc}
 \frac{u_2-u_1}{\beta -u_1+u_2-1} & \frac{\beta -1}{\beta -u_1+u_2-1} \\
 \frac{\beta -1}{\beta -u_1+u_2-1} & \frac{u_2-u_1}{\beta -u_1+u_2-1} \\
\end{array}
\right)
\end{equation}

\subsection{R-Matrix at level 2}
\label{sec:r-matrix-at-1}

\subsubsection{$(q,t)$-deformed version}
\label{sec:q-t-deformed-1}
The generalized Macdonald polynomials at level 2 in the $N=3$ case are expressed as
\begin{align}
&{}^t\left(
\begin{array}{ccccccccc}
 M_{\emptyset,\emptyset,[2]}
& M_{\emptyset,\emptyset,[1,1]}
& M_{\emptyset,[1],[1]}
& M_{[1],\emptyset,[1]}
& M_{\emptyset,[2],\emptyset}
& M_{\emptyset,[1,1],\emptyset}
& M_{[1],[1],\emptyset}
&M_{[2],\emptyset,\emptyset}
&M_{[1,1],\emptyset,\emptyset}
\end{array}
\right)\nonumber\\
&=\mathcal{A} \,
{}^t\left(
\begin{array}{ccccccccc}
 M'_{\emptyset,\emptyset,[2]}
& M'_{\emptyset,\emptyset,[1,1]}
& M'_{\emptyset,[1],[1]}
& M'_{[1],\emptyset,[1]}
& M'_{\emptyset,[2],\emptyset}
& M'_{\emptyset,[1,1],\emptyset}
& M'_{[1],[1],\emptyset}
& M'_{[2],\emptyset,\emptyset}
& M'_{[1,1],\emptyset,\emptyset}
\end{array}
\right),
\end{align}
where $M'_{ABC}$ denotes the product of ordinary Macdonald polynomials $M_A(p^{(1)})M_B(p^{(2)})M_C(p^{(3)})$,
and the matrix $\mathcal{A}$ is given below.
In the same manner, one can get the representation matrix of $\mathcal{R}$.
First of all, we choose $B^{(12)}$ at level 2 to be of the form
\begin{equation}
\widetilde{B}^{(12)}=
\left(
\begin{array}{ccccccccc}
1& 0 & 0 & 0 & 0 & 0 & 0 & 0 & 0 \\
0& 1 & 0 & 0 & 0 & 0 & 0 & 0 & 0 \\
b_{31}& b_{32} & * & * & 0 & 0 & 0 & 0 & 0 \\
b_{41}& b_{42} & * & * & 0 & 0 & 0 & 0 & 0 \\
b_{51}& b_{52} & b_{53} & b_{54} & * & * & * & * & * \\
b_{61}& b_{62} & b_{63} & b_{64} & * & * & * & * & * \\
b_{71}& b_{72} & b_{73} & b_{74} & * & * & * & * & * \\
b_{81}& b_{82} & b_{83} & b_{84} & * & * & * & * & * \\
b_{91}& b_{92} & b_{93} & b_{94} & * & * & * & * & * \\
\end{array}
\right).
\end{equation}
Then, one finds the proportionality constant such that all $b_{ij}$ are
zero just by solving the equations $b_{i1}=0$ ($i=3,4,\ldots ,9$). We also checked that the representation matrix $\widetilde{B}^{ij}$ obtained in this way satisfies the Yang-Baxter equation.

\begin{landscape}

Examples of the generalized Macdonald polynomials.
\begin{equation}
\mathcal{A}=
\left(
\begin{array}{ccccccccc}
 1 & 0 & -\frac{q (q+1) (q-t) (t-1) u_3}{\sqrt{\frac{q}{t}} t (q t-1) (u_2-q
   u_3)} & \frac{(q+1) (t-1) (t-q) u_3 \left(t u_2-q^2 u_3\right)}{t (q t-1)
   (q u_3-u_1) (q u_3-u_2)} & -\frac{(q-t) u_3 \left(t u_3 q^3-t u_2 q^2+t
   u_3 q^2-u_3 q^2-t^2 u_3 q+t u_2\right)}{q t (q t-1) (u_2-u_3) (q
   u_3-u_2)} & -\frac{(q-1) (q+1) (q-t) (t-1) (t+1) u_3}{(q t-1)^2 (q
   u_3-u_2)} \\
 0 & 1 & -\frac{(q-t) u_3}{\sqrt{\frac{q}{t}} t (t u_2-u_3)} & \frac{(q-t) u_3
   \left(q u_3-t^2 u_2\right)}{q t (t u_1-u_3) (t u_2-u_3)} & \frac{(q-t)
   u_3}{q (t u_2-u_3)} & \frac{(q-t) u_3 \left(-q u_2 t^3+u_3 t^2+q u_2
   t+q^2 u_3 t-q u_3 t-q u_3\right)}{q t (q t-1) (u_2-u_3) (t u_2-u_3)} \\
 0 & 0 & 1 & -\frac{(q-t) \sqrt{\frac{q}{t}} u_2}{q (u_1-u_2)} & -\frac{(q-t)
   u_3}{\sqrt{\frac{q}{t}} t (q u_2-u_3)} & -\frac{(q-1) (q-t) (t+1)
   u_3}{\sqrt{\frac{q}{t}} (q t-1) (u_2-t u_3)} \\
 0 & 0 & 0 & 1 & 0 & 0 \\
 0 & 0 & 0 & 0 & 1 & 0 \\
 0 & 0 & 0 & 0 & 0 & 1 \\
 0 & 0 & 0 & 0 & 0 & 0 \\
 0 & 0 & 0 & 0 & 0 & 0 \\
 0 & 0 & 0 & 0 & 0 & 0 \\
\end{array}
\right.
\end{equation}
\begin{equation}
\left.
\begin{array}{ccccccccc}
  -\frac{(q+1) (q-t)^2 (t-1) u_3^2 \left(q^2 u_3-t
   u_2\right)}{\sqrt{\frac{q}{t}} t^2 (q t-1) (u_2-u_3) (u_1-q u_3) (u_2-q
   u_3)} & -\frac{(q-t) u_3 (q u_3-t u_2) \left(q^2 u_3-t u_2\right) \left(t
   u_3 q^3-t u_1 q^2+t u_3 q^2-u_3 q^2-t^2 u_3 q+t u_1\right)}{q^2 t^2 (q
   t-1) (u_1-u_3) (u_3-u_2) (q u_3-u_1) (q u_3-u_2)} & \frac{(q-1) (q+1)
   (q-t) (t-1) (t+1) u_3 (q u_3-t u_2) \left(q^2 u_3-t u_2\right)}{q t (q
   t-1)^2 (u_2-u_3) (q u_3-u_1) (q u_3-u_2)} \\
   -\frac{(q-t)^2 \sqrt{\frac{q}{t}} u_3^2 \left(q u_3-t^2 u_2\right)}{q^2 t (t
   u_1-u_3) (u_2-u_3) (t u_2-u_3)} & -\frac{(q-t) u_3 (q u_3-t u_2)
   \left(q u_3-t^2 u_2\right)}{q^2 t (t u_1-u_3) (t u_2-u_3) (u_3-u_2)} &
   -\frac{(q-t) u_3 (q u_3-t u_2) \left(q u_3-t^2 u_2\right) \left(-q u_1
   t^3+u_3 t^2+q u_1 t+q^2 u_3 t-q u_3 t-q u_3\right)}{q^2 t^2 (q t-1)
   (u_1-u_3) (t u_1-u_3) (t u_2-u_3) (u_3-u_2)} \\
 a_{37} & a_{38} & a_{39} \\
 -\frac{(q-t) u_3}{\sqrt{\frac{q}{t}} t (u_2-u_3)} &
   -\frac{(q-t) u_3 (q u_3-t u_2)}{q t (q u_1-u_3) (u_3-u_2)} &
   -\frac{(q-1) (q-t) (t+1) u_3 (q u_3-t u_2)}{q (q t-1) (u_2-u_3) (t
   u_3-u_1)} \\
 -\frac{q (q+1) (q-t) (t-1) u_2}{\sqrt{\frac{q}{t}} t (q
   t-1) (u_1-q u_2)} & -\frac{(q-t) u_2 \left(t u_2 q^3-t u_1 q^2+t u_2
   q^2-u_2 q^2-t^2 u_2 q+t u_1\right)}{q t (q t-1) (u_1-u_2) (q u_2-u_1)} &
   -\frac{(q-1) (q+1) (q-t) (t-1) (t+1) u_2}{(q t-1)^2 (q u_2-u_1)} \\
 -\frac{(q-t) u_2}{\sqrt{\frac{q}{t}} t (t u_1-u_2)} &
   \frac{(q-t) u_2}{q (t u_1-u_2)} & \frac{(q-t) u_2 \left(-q u_1 t^3+u_2
   t^2+q u_1 t+q^2 u_2 t-q u_2 t-q u_2\right)}{q t (q t-1) (u_1-u_2) (t
   u_1-u_2)} \\
 1 & -\frac{(q-t) \sqrt{\frac{q}{t}} u_2}{q (q
   u_1-u_2)} & -\frac{(q-1) (q-t) \sqrt{\frac{q}{t}} t (t+1) u_2}{q (q t-1)
   (u_1-t u_2)} \\
 0 & 1 & 0 \\
 0 & 0 & 1 \\
\end{array}
\right)\nn
\end{equation}
\begin{align}
&a_{37}=\frac{u_3 (q-t) \left(q^2 u_2 ((t+1) u_2 u_3-u_1 (u_2+u_3))+q \left(t^2
   (-u_1) u_2 u_3+t \left(u_1 \left(2 u_2^2+2 u_3 u_2+u_3^2\right)-u_2
   \left(u_2^2+2 u_3 u_2+2 u_3^2\right)\right)+u_2^2 u_3\right)+t u_2 u_3
   (t (u_2+u_3)-(t+1) u_1)\right)}{q t (u_1-u_2) (u_1-u_3) (q u_2-u_3) (t
   u_3-u_2)}\\
&a_{38}=-\frac{(q-t)^2 u_2 u_3 \left(u_2 u_3 t^2-q
   u_2^2 t-q u_3^2 t+q u_1 u_2 t-u_1 u_2 t+q u_1 u_3 t-u_1 u_3 t-q u_2
   u_3 t+u_2 u_3 t+q u_2 u_3\right)}{q \sqrt{\frac{q}{t}} t^2 (u_1-u_2)
   (u_1-u_3) (q u_2-u_3) (u_2-t u_3)}\\
&a_{39}=-\frac{(q-1) (q-t)^2 (t+1) u_2 u_3
   \left(u_2 u_3 q^2-u_2^2 q-u_3^2 q-t u_1 u_2 q+u_1 u_2 q-t u_1 u_3
   q+u_1 u_3 q+t u_2 u_3 q-u_2 u_3 q+t u_2 u_3\right)}{q
   \sqrt{\frac{q}{t}} t (q t-1) (u_1-u_2) (u_1-u_3) (q u_2-u_3) (u_2-t
   u_3)}
\end{align}

The representation matrix of $\mathcal{R}$ in the basis of generalized Macdonald polynomials
\begin{align}
&\left(
\begin{array}{ccccc}
 -\frac{(q-Q t) \left(q^2-Q t\right)}{q (q-Q) (Q-1) t} & 0 & \frac{Q (q-t)
   \sqrt{\frac{q}{t}} (Q t-1)}{q (q Q-1) (Q-t)} & \\
 0 & \frac{(q-Q t) \left(q-Q t^2\right)}{q (Q-1) t (Q t-1)} & -\frac{(q-1) (q-Q) Q
   (q-t) \sqrt{\frac{q}{t}} t (t+1)}{q (q Q-1) (Q-t) (q t-1)}  \\
 \frac{(q+1) (q-t) (t-1) (q-Q t) \left(q^2-Q t\right)}{(q-Q)^2 (Q-1)
   \sqrt{\frac{q}{t}} t^2 (q t-1)} & \frac{(q-t) (q-Q t) \left(q-Q t^2\right)}{q
   (Q-1) \sqrt{\frac{q}{t}} t^2 (Q t-1)^2} &v_{33}& \\
 -\frac{(q-t) (q-Q t) \left(q^2-Q t\right) \left(Q t q^3-q^2-t^2 q-Q t q+t
   q+t\right)}{q^2 (q-Q) (Q-1)^2 (q Q-1) t^2 (q t-1)} & -\frac{(q-t) (q-Q t)
   \left(q-Q t^2\right)}{q (Q-1) (q Q-1) t^2 (Q t-1)} & v_{43} \\
 -\frac{(q-1) (q+1) (q-t) (t-1) (t+1) (q-Q t) \left(q^2-Q t\right)}{q (q-Q) (Q-1)
   t (t-Q) (q t-1)^2} & -\frac{(q-t) (q-Q t) \left(q-Q t^2\right) \left(-q t^3-q
   t^2+q Q t^2+t^2+q^2 t-q Q\right)}{q^2 (Q-1)^2 t^2 (t-Q) (q t-1) (Q t-1)} &
  v_{53} \\
\end{array}
\right. \\
& \qquad
\left.
\begin{array}{ccccc}
\frac{Q (q-t) \left(Q t q^3-Q
   q^2+Q t q^2-t q^2-Q t^2 q+t\right)}{(q Q-t) \left(q^2 Q-t\right) (q t-1)} &
   -\frac{(Q-1) Q (q-t) t}{(q Q-t) \left(q Q-t^2\right)} \\
 -\frac{(q-1) q
   (q+1) (Q-1) Q (q-t) (t-1) t (t+1)}{(q Q-t) \left(q^2 Q-t\right) (q t-1)^2} &
   \frac{Q (q-t) \left(-q t^3+Q t^2+q t+q^2 Q t-q Q t-q Q\right)}{(q Q-t) (q t-1)
   \left(q Q-t^2\right)} \\
 v_{34} & v_{35} \\
   v_{44}& -\frac{Q
   (q-t)^2 \left(Q q^2+Q t q^2-t^2 q-Q t q+t q-t^2\right)}{q (q Q-1) (q Q-t) t
   \left(q Q-t^2\right)} \\
 -\frac{(q-1) (q+1) Q (q-t)^2 (t-1) (t+1) \left(Q q^2+Q q-Q t q+t
   q-t^2-t\right)}{(Q-t) (q Q-t) \left(q^2 Q-t\right) (q t-1)^2} &
v_{55} \\
\end{array}
\right)\nn
\end{align}
\begin{align}
&\textstyle v_{33}= \frac{Q q^4+Q t q^4+Q^2 t^3 q^3+Q^3
   t^2 q^3-4 Q t^2 q^3+Q q^3-Q^3 t q^3-4 Q^2 t q^3-3 Q t q^3+t q^3-4 Q^3 t^3 q^2+Q
   t^3 q^2+3 Q^3 t^2 q^2}{q (q-Q) (q Q-1) t (t-Q) (Q t-1)} \\
&\textstyle \qquad +\frac{12 Q^2 t^2 q^2+3 Q t^2 q^2+Q^3 t q^2-4 Q t q^2+Q^3 t^4
   q+Q^4 t^3 q-3 Q^3 t^3 q-4 Q^2 t^3 q-Q t^3 q-4 Q^3 t^2 q+Q t^2 q+Q^2 t q+Q^3
   t^4+Q^3 t^3}{q (q-Q) (q Q-1) t (t-Q) (Q t-1)} \\
&\textstyle v_{43}=\frac{(q-t)
   \sqrt{\frac{q}{t}} \left(-Q q^3+Q^3 t q^3+Q^2 t q^3-Q t q^3-Q^3 t^2 q^2-2 Q^2
   t^2 q^2+2 Q t^2 q^2-Q^3 t q^2+3 Q t q^2-t q^2+2 Q^3 t^2 q-2 Q^2 t^2 q-Q t^2
   q-Q^2 t q+2 Q t q+Q^2 t^3-Q t^2\right)}{q^2 (Q-1) (q Q-1)^2 (Q-t) t}\\
&\textstyle v_{53}=-\frac{(q-1) (t+1) (q-t) \sqrt{\frac{q}{t}} \left(q^3 Q+q^2 \left(Q^3 (t-1)+Q^2
   \left(t^2-1\right)+Q \left(-2 t^2-3 t+1\right)+t\right)+q Q t \left(Q^2 (1-2
   t)+2 Q (t+1)+t^2+t-2\right)-Q^2 t^3\right)}{q^2 (Q-1) (q Q-1) (q t-1) (Q-t)^2}\\
&\textstyle v_{44}=\frac{q^7 Q^3 t-q^6 Q^2 \left(2 Q t^2-Q t+t+1\right)+q^5 Q t \left(Q^3 t^2-Q^2
   \left(t^2+t+2\right)+Q \left(t^2-3 t+5\right)+2 t-1\right)-q^4 Q t \left(Q^3
   t+Q^2 \left(3 t^2-7 t+2\right)+Q \left(-13 t^2+11 t-5\right)\right)}
{q (Q-1) t (q Q-1) (q t-1) (q Q-t) \left(q^2Q-t\right)} \\
&\textstyle \qquad +\frac{-q^4 Q t(6 t^2-4t+1)+q^3 t^2 \left(Q^3 \left(t^2-4 t+6\right)+Q^2 \left(-5 t^2+11
   t-13\right)+Q \left(2 t^2-7 t+3\right)+t\right)
+q Q t^3 \left(Qt^2+(Q-1) t+2\right)-Q t^4 +q^2 t^2 \left(Q^3 (t-2) t+Q^2
   \left(-5 t^2+3 t-1\right)+Q \left(2 t^2+t+1\right)-1\right)}
{q (Q-1) t (q Q-1) (q t-1) (q Q-t) \left(q^2Q-t\right)} \\
&\textstyle v_{34}=-\frac{q (q+1) Q (q-t) (t-1)
   \left(-Q q^3+Q^2 t q^3-2 Q^2 t^2 q^2+Q t^2 q^2+Q^2 t q^2+2 Q t q^2-2 t q^2+Q^3
   t^2 q-3 Q^2 t^2 q+t^2 q-2 Q^2 t q+2 Q t q+t q+Q^2 t^3+Q^2 t^2-Q
   t^2-t^2\right)}{(q-Q) (q Q-t) \left(q^2 Q-t\right) \sqrt{\frac{q}{t}} t (q t-1)
   (Q t-1)}\\
&\textstyle v_{35}=\frac{Q (q-t) \left(Q q^3-Q^2 q^2+2 Q^2 t^2 q^2-2 Q t^2 q^2-t^2
   q^2-Q^2 t q^2-2 Q t q^2+2 t q^2-Q^2 t^3 q+Q t^3 q+t^3 q-Q^3 t^2 q+3 Q^2 t^2
   q-t^2 q+2 Q^2 t q-Q t q-Q^2 t^3\right)}{(q-Q) (q Q-t) \sqrt{\frac{q}{t}} t (Q
   t-1) \left(q Q-t^2\right)}\\
&\textstyle v_{55}=\frac{q^5 Q^2 t+q^4 Q \left(Q^2 \left(2 t^3+2 t^2-t-1\right)+Q \left(-5 t^3-5
   t^2+t\right)+t^2 (t+1)\right)+q^3 t \left(Q^4 t^2+Q^3 \left(-6 t^3-7
   t^2+t+2\right)+Q^2 t \left(t^3+13 t^2+11 t+3\right)-Q t \left(t^3+3 t^2+4
   t+2\right)+t^4\right)}{q(Q-1) t (q t-1) (Q-t) (q Q-t) \left(q Q-t^2\right)}\\
& \textstyle \qquad +\frac{-q^2 t^2 \left(Q^4-Q^3 \left(2 t^3+4 t^2+3 t+1\right)+Q^2
   \left(3 t^3+11 t^2+13 t+1\right)+Q t \left(2 t^3+t^2-7 t-6\right)+t^2\right)-q
   Q t^4 \left(Q^2 (t+1)+Q \left(t^2-5 t-5\right)-t^3-t^2+2 t+2\right)-Q^2 t^6}
{q(Q-1) t (q t-1) (Q-t) (q Q-t) \left(q Q-t^2\right)}
\end{align}

The representation matrix of $\mathcal{R}$ in the basis of power sum symmetric functions,
\begin{equation}
\left(
\begin{array}{ccccc}
r_{11} & -\frac{(q-1) q
   (Q-1) Q (q-t) t (t+1)}{2 (q Q-t) \left(q^2 Q-t\right) \left(q Q-t^2\right)} &
   \frac{(q-1) (Q-1) Q (q-t) \sqrt{\frac{q}{t}} t^2 (t+1)}{2 (q Q-t) \left(q^2
   Q-t\right) \left(q Q-t^2\right)} &
r_{14}&
   -\frac{(q-1) (Q-1) Q (q-t) t^2 (t+1)}{2 (q Q-t) \left(q^2 Q-t\right) \left(q
   Q-t^2\right)} \\
 -\frac{q (q+1) (Q-1) Q (q-t) (t-1) t}{2 (q Q-t) \left(q^2 Q-t\right) \left(q
   Q-t^2\right)} & r_{22}& \frac{(Q-1) Q (q-t) \sqrt{\frac{q}{t}} t \left(2 Q q^2-t^2 q+t
   q-t^2-t\right)}{2 (q Q-t) \left(q^2 Q-t\right) \left(q Q-t^2\right)} &
   \frac{(q+1) (Q-1) Q (q-t) (t-1) t^2}{2 (q Q-t) \left(q^2 Q-t\right) \left(q
   Q-t^2\right)} & r_{25} \\
 \frac{q^2 (q+1) (Q-1) Q (q-t) (t-1)}{(q Q-t) \left(q^2 Q-t\right)
   \sqrt{\frac{q}{t}} \left(q Q-t^2\right)} & \frac{q (Q-1) (q-t) \left(Q q^2+Q t
   q^2+Q q-Q t q-2 t^2\right)}{(q Q-t) \left(q^2 Q-t\right) \sqrt{\frac{q}{t}}
   \left(q Q-t^2\right)} & r_{33}& -\frac{q
   (q+1) (Q-1) Q (q-t) (t-1) t}{(q Q-t) \left(q^2 Q-t\right) \sqrt{\frac{q}{t}}
   \left(q Q-t^2\right)} & \frac{q (Q-1) Q (q-t) \left(2 Q q^2-t^2 q+t
   q-t^2-t\right)}{(q Q-t) \left(q^2 Q-t\right) \sqrt{\frac{q}{t}} \left(q
   Q-t^2\right)} \\
r_{41}& \frac{(q-1) q^2 (Q-1) Q (q-t) (t+1)}{2 (q
   Q-t) \left(q^2 Q-t\right) \left(q Q-t^2\right)} & -\frac{(q-1) q (Q-1) Q (q-t)
   \sqrt{\frac{q}{t}} t (t+1)}{2 (q Q-t) \left(q^2 Q-t\right) \left(q
   Q-t^2\right)} &
r_{44}& \frac{(q-1) q (Q-1) Q (q-t) t (t+1)}{2 (q Q-t) \left(q^2
   Q-t\right) \left(q Q-t^2\right)} \\
 -\frac{q^2 (q+1) (Q-1) Q (q-t) (t-1)}{2 (q Q-t) \left(q^2 Q-t\right) \left(q
   Q-t^2\right)} &
 r_{52}& \frac{(Q-1) (q-t)
   \sqrt{\frac{q}{t}} t \left(Q q^2+Q t q^2+Q q-Q t q-2 t^2\right)}{2 (q Q-t)
   \left(q^2 Q-t\right) \left(q Q-t^2\right)} & \frac{q (q+1) (Q-1) Q (q-t) (t-1)
   t}{2 (q Q-t) \left(q^2 Q-t\right) \left(q Q-t^2\right)} & r_{55} \\
\end{array}
\right)
\end{equation}

\begin{align}
&r_{11}= -\frac{q (Q-1) t \left(-2 Q^2 q^2-Q q^2+Q t q^2+Q t^2 q+Q q+2 Q t q-Q t^2-2 t^2+Q
   t\right)}{2 (q Q-t) \left(q^2 Q-t\right) \left(q Q-t^2\right)}, \\
&r_{41}= \frac{(q-t) \left(Q^2 q^3+Q q^3+Q^2 t q^3-Q t q^3+Q^2 q^2-2 Q t^2 q^2-Q q^2+Q^2 t
   q^2-Q t q^2-2 Q t^2 q+2 t^2 q-2 Q t q+2 t^3\right)}{2 (q Q-t) \left(q^2
   Q-t\right) \left(q Q-t^2\right)} \\
&r_{22}=-\frac{q (Q-1) t \left(-2 Q^2 q^2+Q q^2+Q t q^2+Q t^2 q+Q q-2 Q
   t q+Q t^2-2 t^2+Q t\right)}{2 (q Q-t) \left(q^2 Q-t\right) \left(q
   Q-t^2\right)} \\
&r_{52}=-\frac{(q-t) \left(-Q^2 q^3-Q q^3+Q^2 t q^3-Q t q^3+Q^2 q^2+2 Q
   t^2 q^2-Q q^2-Q^2 t q^2+Q t q^2-2 Q t^2 q+2 t^2 q+2 Q t q-2 t^3\right)}{2 (q
   Q-t) \left(q^2 Q-t\right) \left(q Q-t^2\right)} \\
&r_{33}=\frac{Q^2 q^4-Q^2 t^2 q^3+Q^3 t q^3-3 Q^2 t q^3+Q t
   q^3+Q t^3 q^2+2 Q^2 t^2 q^2-2 Q t^2 q^2-Q^2 t q^2-Q^2 t^3 q+3 Q t^3 q-t^3 q+Q
   t^2 q-Q t^4}{(q Q-t) \left(q^2 Q-t\right) \left(q Q-t^2\right)} \\
&r_{14}=\frac{Q (q-t) \left(2 Q^2 q^3-2 Q t^2 q^2+2
   Q^2 t q^2-2 Q t q^2-Q t^3 q+t^3 q-Q t^2 q+t^2 q-2 Q t q+Q t^3+t^3-Q
   t^2+t^2\right)}{2 (q Q-t) \left(q^2 Q-t\right) \left(q Q-t^2\right)} \\
&r_{44}=-\frac{q (Q-1) t \left(-2 Q^2 q^2-Q q^2+Q t q^2+Q t^2 q+Q q+2 Q
   t q-Q t^2-2 t^2+Q t\right)}{2 (q Q-t) \left(q^2 Q-t\right) \left(q
   Q-t^2\right)} \\
&r_{25}=\frac{Q (q-t) \left(2 Q^2 q^3-2 Q t^2 q^2-2 Q^2 t q^2+2 Q t
   q^2+Q t^3 q-t^3 q-Q t^2 q+t^2 q-2 Q t q+Q t^3+t^3+Q t^2-t^2\right)}{2 (q Q-t)
   \left(q^2 Q-t\right) \left(q Q-t^2\right)} \\
&r_{55}=-\frac{q (Q-1) t
   \left(-2 Q^2 q^2+Q q^2+Q t q^2+Q t^2 q+Q q-2 Q t q+Q t^2-2 t^2+Q t\right)}{2 (qQ-t) \left(q^2 Q-t\right) \left(q Q-t^2\right)}
\end{align}
where $Q=\frac{u_1}{u_2}$.

\subsubsection{$\beta$-deformed version}
\label{sec:beta-deform-vers-1}
The representation matrix of $\mathcal{R}^{(\beta)}$ in the basis of generalized Jack polynomials
\begin{align}
\mathcal{R}^{(\beta)}=&\left(
\begin{array}{ccccc}
 \frac{(a+\beta -2) (a+\beta -1)}{(a-1) a} & 0 & -\frac{(\beta -1) (a+\beta
   )}{(a+1) (a-\beta )} \\
 0 & \frac{(a+\beta -1) (a+2 \beta -1)}{a (a+\beta )} & -\frac{2 (a-1) (\beta
   -1)}{(a+1) (a-\beta ) (\beta +1)} \\
 -\frac{2 (\beta -1) \beta  (a+\beta -2) (a+\beta -1)}{(a-1)^2 a (\beta +1)} &
   -\frac{(\beta -1) (a+\beta -1) (a+2 \beta -1)}{a (a+\beta )^2} & s_{33} &
 \\
 -\frac{(\beta -1) (a+\beta -2) (a+\beta -1) \left(-\beta ^2+\beta +2
   a+2\right)}{(a-1) a^2 (a+1) (\beta +1)} & \frac{(\beta -1) (a+\beta -1) (a+2
   \beta -1)}{a (a+1) (a+\beta )} & -\frac{(\beta -1) \left(a^3+\beta
   a^2-a^2+\beta ^2 a-6 \beta  a+4 a+\beta ^3-3 \beta ^2+2 \beta \right)}{a
   (a+1)^2 (a-\beta )} \\
 \frac{4 (\beta -1) \beta  (a+\beta -2) (a+\beta -1)}{(a-1) a (a-\beta ) (\beta
   +1)^2} & -\frac{(\beta -1) (a+\beta -1) (a+2 \beta -1) \left(-2 \beta ^2+2 a
   \beta -\beta +1\right)}{a^2 (a-\beta ) (\beta +1) (a+\beta )} & -\frac{2 (\beta
   -1) \left(a^3+\beta  a^2-a^2+4 \beta ^2 a-6 \beta  a+a-2 \beta ^2+3 \beta
   -1\right)}{a (a+1) (a-\beta )^2 (\beta +1)}  \\
\end{array}
\right.\\
&
\qquad
\left.
\begin{array}{ccccc}
  \frac{(\beta -1) \left(\beta ^2-\beta -2
   a-2\right)}{(\beta +1) (-a+\beta -2) (-a+\beta -1)} & \frac{a (\beta -1)}{(a-2
   \beta +1) (a-\beta +1)} \\
  \frac{4 a (\beta -1) \beta }{(a-\beta +1)
   (a-\beta +2) (\beta +1)^2} & \frac{(\beta -1) \left(2 \beta ^2-2 a \beta +\beta
   -1\right)}{(\beta +1) (-a+\beta -1) (-a+2 \beta -1)} \\
 -\frac{2 (\beta -1) \beta  \left(a^3+\beta
   a^2-a^2+\beta ^2 a-6 \beta  a+4 a+\beta ^3-3 \beta ^2+2 \beta \right)}{(a-1)
   (a-\beta +1) (a-\beta +2) (\beta +1) (a+\beta )} & -\frac{(\beta -1)
   \left(a^3+\beta  a^2-a^2+4 \beta ^2 a-6 \beta  a+a-2 \beta ^2+3 \beta
   -1\right)}{(a-1) (a-2 \beta +1) (a-\beta +1) (a+\beta )} \\
 s_{44}
   & -\frac{(a-3 \beta +3) (\beta -1)^2}{(a+1) (a-2 \beta +1) (a-\beta +1)} \\
 -\frac{4 (\beta -1)^2 \beta
   (-a+3 \beta -3)}{(\beta +1)^2 (-a+\beta -2) (-a+\beta -1) (\beta -a)} & \frac{4
   \beta ^5-8 a \beta ^4-8 \beta ^4+5 a^2 \beta ^3+20 a \beta ^3+\beta ^3-2 a^3
   \beta ^2-5 a^2 \beta ^2-16 a \beta ^2+7 \beta ^2+a^4 \beta -2 a^3 \beta +4 a
   \beta -5 \beta +a^4+2 a^2+1}{a (a-2 \beta +1) (a-\beta ) (a-\beta +1) (\beta
   +1)} \\
\end{array}
\right)\nn
\end{align}

\begin{align}
&s_{33}=\frac{a^4+2
   \beta  a^3-2 a^3+3 \beta ^2 a^2-8 \beta  a^2+3 a^2+4 \beta ^3 a-14 \beta ^2
   a+14 \beta  a-4 a+2 \beta ^4-6 \beta ^3+9 \beta ^2-6 \beta +2}{(a-1) (a+1)
   (a-\beta ) (a+\beta )}\\
&s_{44}=\frac{\beta ^5-5 \beta ^4+2 a^2 \beta ^3-4 a \beta ^3+7
   \beta ^3+16 a \beta ^2+\beta ^2+a^4 \beta +2 a^3 \beta -5 a^2 \beta -20 a \beta
   -8 \beta +a^4+2 a^3+5 a^2+8 a+4}{a (a+1) (a-\beta +1) (a-\beta +2) (\beta +1)}
\end{align}

The representation matrix of $\mathcal{R}^{(\beta)}$ in the basis of power sum symmetric functions
\begin{equation}
\left(
\begin{array}{ccccc}
 \frac{a \left(a^2-2 \beta  a+2 a+\beta ^2-3 \beta +1\right)}{(a-2 \beta +1)
   (a-\beta +1) (a-\beta +2)} & \frac{a (\beta -1)}{(a-2 \beta +1) (a-\beta +1)
   (a-\beta +2)} & -\frac{a (\beta -1)}{(a-2 \beta +1) (a-\beta +1) (a-\beta +2)}
   & \frac{(\beta -1) \left(2 a^2-4 \beta  a+4 a+2 \beta ^2-5 \beta
   +2\right)}{(-a+\beta -2) (-a+\beta -1) (-a+2 \beta -1)} & \frac{a (\beta
   -1)}{(a-2 \beta +1) (a-\beta +1) (a-\beta +2)} \\
 \frac{a (\beta -1) \beta }{(a-2 \beta +1) (a-\beta +1) (a-\beta +2)} & \frac{a
   \left(a^2-2 \beta  a+2 a-\beta \right)}{(a-2 \beta +1) (a-\beta +1) (a-\beta
   +2)} & -\frac{a (a-2 \beta +2) (\beta -1)}{(a-2 \beta +1) (a-\beta +1) (a-\beta
   +2)} & -\frac{a (\beta -1) \beta }{(a-2 \beta +1) (a-\beta +1) (a-\beta +2)} &
   \frac{(\beta -1) \left(2 \beta ^2-a \beta -5 \beta +a+2\right)}{(-a+\beta -2)
   (-a+\beta -1) (-a+2 \beta -1)} \\
 -\frac{2 a (\beta -1) \beta }{(a-2 \beta +1) (a-\beta +1) (a-\beta +2)} &
   -\frac{2 a (a-2 \beta +2) (\beta -1)}{(a-2 \beta +1) (a-\beta +1) (a-\beta +2)}
   & \frac{a^3-2 \beta  a^2+2 a^2+\beta ^2 a-3 \beta  a+a-2 \beta ^3+7 \beta ^2-7
   \beta +2}{(a-2 \beta +1) (a-\beta +1) (a-\beta +2)} & \frac{2 a (\beta -1)
   \beta }{(a-2 \beta +1) (a-\beta +1) (a-\beta +2)} & -\frac{2 a (a-2 \beta +2)
   (\beta -1)}{(a-2 \beta +1) (a-\beta +1) (a-\beta +2)} \\
 \frac{(\beta -1) \left(2 a^2-4 \beta  a+4 a+2 \beta ^2-5 \beta
   +2\right)}{(-a+\beta -2) (-a+\beta -1) (-a+2 \beta -1)} & -\frac{a (\beta
   -1)}{(a-2 \beta +1) (a-\beta +1) (a-\beta +2)} & \frac{a (\beta -1)}{(a-2 \beta
   +1) (a-\beta +1) (a-\beta +2)} & \frac{a \left(a^2-2 \beta  a+2 a+\beta ^2-3
   \beta +1\right)}{(a-2 \beta +1) (a-\beta +1) (a-\beta +2)} & -\frac{a (\beta
   -1)}{(a-2 \beta +1) (a-\beta +1) (a-\beta +2)} \\
 \frac{a (\beta -1) \beta }{(a-2 \beta +1) (a-\beta +1) (a-\beta +2)} &
   \frac{(\beta -1) \left(2 \beta ^2-a \beta -5 \beta +a+2\right)}{(-a+\beta -2)
   (-a+\beta -1) (-a+2 \beta -1)} & -\frac{a (a-2 \beta +2) (\beta -1)}{(a-2 \beta
   +1) (a-\beta +1) (a-\beta +2)} & -\frac{a (\beta -1) \beta }{(a-2 \beta +1)
   (a-\beta +1) (a-\beta +2)} & \frac{a \left(a^2-2 \beta  a+2 a-\beta
   \right)}{(a-2 \beta +1) (a-\beta +1) (a-\beta +2)} \\
\end{array}
\right)
\end{equation}
where $a=u_1-u_2$.
\end{landscape}

\section{Realization of rank $N$ representation by generalized Macdonald polynomials}
\label{sec:high-hamilt-gener}

One can consider a representation of the DIM algebra which is called rank $N$ representation
and can be realized in terms of a basis $\tket{\vu, \vl}$ called AFLT basis, \cite{Bourgine:2016vsq}.
This representation is given by the $N$-fold tensor product of the level (0,1) representations (i.e. the vertical representations in terms of our paper, which are spectral dual to the level (1,0) (horizontal) representations) which are realized by free bosons for the refined topological vertex.
In this Appendix, which is based on the spectral duality,
we present conjectures for explicit expressions
of the action of $x^{+}_{\pm 1}$ on the generalized Macdonald polynomials,
which are defined to be eigenfunctions of the Hamiltonian $\Xo_0$.
We also conjecture the eigenvalues of higher Hamiltonians acting on the
generalized Macdonald polynomials from those of the spectral
dual generators provided in \cite{Bourgine:2016vsq}.
Our conjectures mean that
the generalized Macdonald polynomials explicitly realize
the spectral dual basis to $\tket{\vu, \vl}$ in \cite{Bourgine:2016vsq}.

\subsection{Action of $x_{\pm 1}^{+}$ on generalized Macdonald polynomials}

We use the notation
\begin{align}
X^{(1)}(z) = \sum_{n\in \mathbb{Z}} X^{(1)}_n z^{-n} =
\rho^{(N)}_{u_1,\ldots,u_N} (x^{+}(z)).
\end{align}
For an $N$-tuple of Young diagrams $\vl = (\lo, \ldots \lN)$,
the generalized Macdonald polynomials $M_{\vl}$ are defined to be eigenfunctions
of $\Xo_0$ with the eigenvalues
\begin{equation}
e^{(1)}_{\vl} = \sum_{k=1}^N u_k \left\{ 1+ (t-1)\sum_{i=1}^{\ell(\lambda^{(k)})} (q^{\lambda^{(k)}_i}-1) t^{-i} \right\}.
\end{equation}
$M_{\vl}$ is renormalized as $M_{\vl} = m_{\vl} + \cdots $, in terms of
the product of the monomial symmetric functions $m_{\vl}=m_{\lo} \otimes \cdots \otimes m_{\lambda^{(N)}}$.
Their integral forms $\widetilde{M}_{\vl}$ are defined by
\begin{equation}
\widetilde{M}_{\vl} =  M_{\vl} \times
\prod_{1\leq i<j \leq N} G_{\lambda^{(j)}, \lambda^{(i)}}(u_j/u_i|q,t)
\prod_{k=1}^N \prod_{(i,j)\in \lambda^{(k)}}
(1-q^{\lambda^{(k)}_i-j} t^{\lambda^{(k)\mathrm{T}}_j-i+1}),
\end{equation}
where $\lambda^{\mathrm{T}}$ is the transposed of Young diagram $\lambda$
and we use the Nekrasov factor (\ref{G}).
It is expected that
the basis $\widetilde{M}_{\vl}$ corresponds to
the AFLT basis\footnote{Originally, the AFLT basis is defined by the property that
the inner products and matrix elements of vertex operators
reproduce the Nekrasov factor. In \cite{awata2011notes},
the integral forms $\widetilde{M}_{\vl}$ were already
conjectured for the AFLT basis in this original sense.}
 in \cite{Bourgine:2016vsq}
and realizes the rank $N$ representation through the spectral duality $\cal S$.
That is to say, for any generator $a$ in the DIM algebra, that
the action of $\rho^{(N)}_{u_1,\ldots,u_N} \circ {\cal S}(a)$
on the integral forms $\widetilde{M}_{\vl}$ are
the same as the action of $\rho^{\mathrm{rank} N} (a)$ on the basis $\tket{\vu,\vl}$
\cite{Bourgine:2016vsq}.
Indeed, one can check that
the action of $x^+_{\pm 1}$ on the generalized Macdonald polynomials is
given by the following conjecture.
Let us denote adding a box to or removing it
from the Young diagram $\vl$ through $A(\vl)$ and $R(\vl)$ respectively.
We also use the notation $\chi_{(\ell, i,j)}=u_{\ell} t^{-i+1}q^{j-1}$
for the triple $x=(\ell,i,j)$, where $(i,j) \in \lambda^{(\ell)}$ are the coordinates of the box of the Young diagram $\lambda^{(\ell)}$.
\begin{conj}\label{conj:Action of X}
\begin{equation}
\Xo_1 \tM_{\vl} \overset{?}{=} \sum_{\substack{|\vm| = |\vl|-1 \\ \vl \supset \vm}} \tilde{c}^{(+)}_{\vl,\vm} \tM_{\vm}, \quad
\Xo_{-1} \tM_{\vl} \overset{?}{=} \sum_{\substack{|\vm| = |\vl|+1\\ \vl \subset \vm}} \tilde{c}^{(-)}_{\vl,\vm} \tM_{\vm},
\end{equation}
where
\begin{align}
&\tilde{c}^{(+)}_{\vl,\vm} =
\xi^{(+)}_x
\frac{\prod_{y \in A(\vl)} (1-\chi_x \chi_y^{-1}  (q/t) )}
{\prod_{\substack{y \in R(\vl)\\ y \neq x}} (1-\chi_x \chi_y^{-1})},
\quad x \in \vl \setminus \vm, \\
& \tilde{c}^{(-)}_{\vl,\vm} =
\xi^{(-)}_x
\frac{\prod_{y \in R(\vl)} (1-\chi_y \chi_x^{-1}  (q/t) )}
{\prod_{\substack{y \in A(\vl)\\ y \neq x}} (1-\chi_y \chi_x^{-1})},
\quad x \in \vm \setminus \vl,
\end{align}
and for the triple $(\ell, i,j)$, we put
\begin{equation}
\xi^{(+)}_{(\ell, i,j)} = (-1)^{N+\ell} p^{-\frac{\ell+1}{2}} t^{(N-\ell)i} q^{(\ell-N+1)j} \frac{\prod_{k=1}^{N-\ell}u_{\ell+k}}{u_{\ell}^{N-\ell-1}}, \quad
\xi^{(-)}_{(\ell, i,j)} = (-1)^{\ell} p^{\frac{\ell-1}{2}} t^{(\ell-2)i} q^{(1-\ell)j} \frac{\prod_{k=1}^{\ell -1}u_k}{u_{\ell}^{\ell-2}}.
\end{equation}
\end{conj}

The actions of $\Xo_{\pm 1}$  in this conjecture come from the corresponding actions of
the generators $f_1$ and $e_1$ in \cite{Bourgine:2016vsq} respectively, i.e., those of
$x^{-}_1$ and $x^+_{1}$ in our notation, which are the spectral duals of $x^{+}_1$ and $x^{+}_{-1}$.
Incidentally, introducing the coefficients
$c^{(\pm)}_{\vl,\vm}=c^{(\pm)}_{\vl,\vm}(q,t|u_1,\ldots,u_N)$ by
\begin{equation}
c^{(\pm)}_{\vl,\vm}=
\prod_{1\leq i<j\leq N} \frac{G_{\mu^{(j)},\mu^{(i)}}(u_j/u_i|q,t)}{G_{\lambda^{(j)},\lambda^{(i)}}(u_j/u_i|q,t)}
\prod_{k=1}^N
\frac{ \prod_{(i,j)\in \mu^{(k)}}
(1-q^{\mu^{(k)}_i-j} t^{\mu^{(k)\mathrm{T}}_j-i+1})}
{ \prod_{(i,j)\in \lambda^{(k)}}
(1-q^{\lambda^{(k)}_i-j} t^{\lambda^{(k)\mathrm{T}}_j-i+1})}
\times
\tilde{c}^{(\pm)}_{\vl,\vm},
\end{equation}
i.e. $\Xo_{\pm 1} M_{\vl}
= \sum_{\vm} c^{(\pm)}_{\vl,\vm} M_{\vm}$,
we can further conjecture that
\begin{equation}\label{eq:rel between cplus and cminus}
c^{(+)}_{\vl,\vm}(q,t|u_1,\ldots , u_N)
\overset{?}{=}
-c^{(-)}_{(\mu^{(N)\mathrm{T}},\ldots,\mu^{(1)\mathrm{T}}),(\lambda^{(N)\mathrm{T}},\ldots,\lambda^{(1)\mathrm{T}})}
(t^{-1},q^{-1}|p^{(N-1)/2}u_N,\ldots,p^{(N-1)/2}u_1).
\end{equation}
We have checked conjecture \ref{conj:Action of X} with respect to $X^{(1)}_1$ and
formula (\ref{eq:rel between cplus and cminus})
with the computer for $|\vl| \leq 5$ for $N=1$,
for $|\vl| \leq 3$ for $N=2, 3$ and
for $|\vl| \leq 2$ for $N=4$.
This conjecture \ref{conj:Action of X} with respect to $\Xo_{-1}$ has been also checked
for the same sizes of $\vm$.

\subsection{Higher Hamiltonians}
For each integer $k\geq 1$,
the spectral dual of $\psi^+_k$ is $H_k$ defined by $H_1 = \Xo_0$ and
\begin{equation}
H_k=[\Xo_{-1},\underbrace{[\Xo_0,\cdots ,[\Xo_0}_{k-2},\Xo_{1}]\cdots]],
\qquad k \geq 2.
\end{equation}
According to \cite{Miki}, $H_k$ are spectral dual to $\psi^+_{k}$ and
consequently mutually commuting: $[H_k,H_l]=0$.
Thus, the generalized Macdonald polynomials $M_{\vl}$ are automatically
eigenfunctions of all $H_k$,
i.e.
$H_k M_{\vl} = e^{(k)}_{\vl} M_{\vl}$.
and $H_k$ can be regarded as higher Hamiltonians
for the generalized Macdonalds polynomials.
Since $H_k$ are the spectral duals to $\psi^+_{k}$, $H_k = {\cal S}(\psi^+_{k})$,
their eigenvalues are expected to be
\begin{conj}
\begin{equation}\label{eq:higher eigenvlue}
e^{(k)}_{\vl} \overset{?}{=}
\frac{(1-q)^{k-1}(1-t^{-1})^{k-1}}{1-p^{-1}}
\oint \frac{dz}{2 \pi \sqrt{-1}z }
\prod_{i=1}^N B^{+}_{\lambda^{(i)}}(u_i z)z^{-k},
\end{equation}
where for the partition $\lambda$ we define
\begin{equation}
B^{+}_{\lambda}(z) =
\frac{1-q^{\lambda_1-1}tz}{1-q^{\lambda_1}z}
\prod_{i=1}^{\infty} \frac{(1-q^{\lambda_{i}}t^{-i} z)(1-q^{\lambda_{i+1}-1}t^{-i+1} z)}{(1-q^{\lambda_{i+1}}t^{-i} z)(1-q^{\lambda_{i}-1}t^{-i+1} z)}.
\end{equation}
\end{conj}
The eigenvalues $e^{(k)}_{\vl}$ correspond to those of the rank $N$ representation of the generators $\psi^+_{k}$
in \cite{Bourgine:2016vsq}.
In the $k=1$ case, the conjecture (\ref{eq:higher eigenvlue}) can be proven.
We have checked it
for $|\vl| \leq 5$ for $N=1$,
for $|\vl| \leq 3$ for $N=2$,
for $|\vl| \leq 2$ for $N=3$ and
for $|\vl| \leq 1$ for $N=4$ in the $k \leq 5$ case.

\section*{Acknowledgements}

We appreciate communications on the subject with Y.~Matsuo, T.~Kimura
and especially with A.~Smirnov.

Our work is supported in part by Grant-in-Aid for Scientific Research (\# 24540210) (H.A.),
(\# 15H05738) (H.K.), for JSPS Fellow (\# 26-10187) (Y.O.) and JSPS Bilateral Joint Projects (JSPS-RFBR collaboration)
\lq\lq Exploration of Quantum Geometry via Symmetry and Duality\rq\rq\
from MEXT, Japan. It is also partly supported by grants 15-31-20832-Mol-a-ved (A.Mor.),
15-31-20484-Mol-a-ved (Y.Z.), 16-32-60047-Mol-a-dk (And.Mor), by RFBR grants 16-01-00291 (A.Mir.),
16-02-01021 (A.Mor.\ and Y.Z.) and 15-01-09242 (An.Mor.), by joint grants 15-51-50034-YaF,
15-51-52031-NSC-a, 16-51-53034-GFEN, 16-51-45029-IND-a.

\end{document}